\documentclass[12pt]{iopart}

\expandafter\let\csname equation*\endcsname\relax
\expandafter\let\csname endequation*\endcsname\relax
\usepackage{amsmath}
\usepackage{wasysym}
\usepackage{graphicx,color}
\usepackage[backend=bibtex,sorting=none]{biblatex}
\usepackage{bm}

\newcommand{\exgra}{{\rmfamily\itshape\mdseries e.g.}}
\newcommand{\idest}{{\rmfamily\itshape\mdseries i.e.}}
\newcommand\orbave[1]{\overline{#1}^{\rm{(O)}}}
\begin{document}
\title{Nonlinear equilibria and transport processes in burning plasmas}
\author{Matteo Valerio Falessi$^{1,2}$, Liu Chen$^{3,4,1}$, Zhiyong Qiu$^{3,1}$ and Fulvio Zonca$^{1,3}$}
\address{$^1$Center for Nonlinear Plasma Science and C.R. ENEA Frascati, C.P. 65, 00044 Frascati, Italy \newline 
$^2$Istituto Nazionale di Fisica Nucleare (INFN), Sezione di Roma, Piazzale Aldo Moro 2, Roma, Italy \newline
$^3$Institute of Fusion Theory and Simulation, School of Physics, Zhejiang University, Hangzhou, China \newline
$^4$Department of Physics and Astronomy, University of California, Irvine CA 92697-4575, USA}
\ead{matteo.falessi@enea.it}
\vspace{10pt}

\begin{abstract}
In this work, we put forward a general phase-space transport theory in axisymmetric tokamak plasmas based upon the concept of zonal state (ZS). Within this theoretical framework, the ZS corresponds to a renormalized plasma nonlinear equilibrium consisting of phase-space zonal structures (PSZS) and zonal electromagnetic fields (ZFs) which evolve self-consistently with symmetry breaking fluctuations and sources/collisions. More specifically, our approach involves deriving governing equations for the evolution of particle distribution functions (i.e, PSZS), which can be used to compute the corresponding macro-/meso-scale evolving magnetized plasma equilibrium adopting the Chew Goldberger Low (CGL) description, separating the spatiotemporal microscale structures. The nonlinear physics of ZFs and of geodesic acoustic modes/energetic particle driven geodesic acoustic modes is then analyzed to illustrate the implications of our theory.
\end{abstract}

\label{sec:org6847c5e}
\section{Introduction}
\label{sec:org8d94d85}
One fundamental  aspect of magnetized fusion plasmas is the study of physics processes that are occurring in burning conditions, where $\alpha$ particles are produced, \exgra, by deuterium-tritium fusion reactions. Understanding the behavior of $\alpha$ particles and, more generally, of energetic particles (EPs)\footnote{In a broader sense, when we refer to energetic particles (EPs), we mean not only charged fusion products, but also suprathermal ions and electrons generated by external power sources for plasma heating and current drive.} in fusion plasmas is crucial, as they play a critical role in mediating couplings between microscopic and macroscopic plasma dynamics significantly influencing its behavior as a complex system \cite{zonca2015nonlinear,chen2016physics,zonca2014energetic}. Reactor relevant plasmas will be fundamentally different with respect to those existing in present day experimental devices due to EP characteristic orbit size and predominant contribution to reactor power balance.

A thorough description of EP transport processes is essential since they involve resonantly excited fluctuations, which have different time scales and structures with respect to thermal plasma instabilities. EPs may induce non-local (global) behaviors as well as excite singular radial mode structures at shear Alfvén wave continuum resonances which, through mode conversion, generate radially propagating microscopic fluctuations that may be absorbed at different radial locations. For these reasons, a global approach is necessary, where the entire plasma volume is described taking into account realistic magnetic geometry and non-uniformities. Due to the  characteristic features of EP distribution functions in the velocity space, the excitation of collective fluctuations around the cyclotron frequency is usually of minor importance \cite{chen2016physics} and, therefore, the relevant resonant frequencies are those characterizing the particle guiding center motion in the equilibrium magnetic fields. Consequently, current research on EP-driven instabilities and transport is generally pursued using nonlinear gyrokinetic theory \cite{frieman1982nonlinear,brizard2007foundations,sugama2000gyrokinetic}. Numerical  simulations of EP physics usually require costly and time-consuming global gyrokinetic codes.  While these simulations provide valuable insight into fundamental physics processes, they typically cover only a relatively limited time range of the dynamics. Thus, they typically do not properly address the coupling of different spatio-temporal scales and, hence, have limited predictive capability of transport processes.

To overcome some of the challenges faced in the description of multi-scale burning plasmas dynamics, we have developed the transport theory of phase-space zonal structures (PSZS) \cite{falessi2019,zonca_2021}. PSZS represent slowly evolving (on the transport time scale) structures in the phase-space that are not affected by fast collisionless dissipation and provide a proper definition of plasma nonlinear equilibrium distribution function extending the concept of plasma transport processes to the phase-space \cite{falessi2019}. This is particularly relevant for weakly collisional plasmas that exhibit significant deviations from local thermodynamic equilibrium, often described by Maxwellian distribution functions.  Notably, using the PSZS theory, the usual plasma transport equations can be obtained as a particular limiting case where the deviation from the local Maxwellian is small. This was demonstrated in a previous work \cite{falessi2019} for the energy and density transport equations. However, in the most general case,  PSZS will not be associated with a reference Maxwellian since they will be the result of the competition between resonantly induced nonlinear transport, sources and only weakly collisional effects, thus requiring a phase-space description. The consistency of the PSZS theory with ``gyrokinetic transport theory'' \cite{parra_2008,abel2013multiscale,sugama_2017} and global gyrokinetic codes stems from its foundation in the well-established nonlinear gyrokinetics \cite{frieman1982nonlinear,brizard2007foundations}, as emphasized in \cite{falessi2019}. The novelty, however, stands in explicitly identifying the part of the toroidally symmetric distribution function that must be incorporated in the non-Maxwellian reference state. Over time, in fact, this contribution may increasingly deviate from the reference equilibrium due to nonlinear processes eventually invalidating the usual transport analyses that rely on local Maxwellian equilibria, for instance. This is evidently the case of transport processes of weakly collisional EPs. For example, the role of PSZS in EP transport due to energetic particle modes (EPM) \cite{Chen_1994} was recently investigated by means of gyrokinetic particle-in-cell simulations \cite{wang_ppcf_2023}. There,  the linear scaling of the chirping rate with mode amplitude of nonlinear coherent EPM fluctuations is numerically demonstrated, consistent with theoretical analyses, which predict the ballistic propagation of the PSZS \cite{zonca2015nonlinear,chen2016physics,zoncarmpp2021}. However, one may argue that even the thermal component of magnetized plasmas, for which Coulomb collision tend to continuously restore the nearly Maxwellian local thermodynamic equilibrium, may have significantly different evolutions on the long time scales if the small but finite deviation from local Maxwellian is accounted for.  In other words, the thermal plasma may evolve into nearly Maxwellian equilibria with completely different radial profiles (cf., \exgra, the recent work in Ref. \cite{wang2023self}). 

In this work, we first derive the PSZS evolution equation in conservative form using the equilibrium constants of motion as phase-space coordinates. The orbit averaging approach, adopted here, has analogies with the methodologies that are used for neoclassical transport studies in stellarators in the weakly collisional regimes \cite{calvo_private}; and in Hamiltonian formulations of quasi-linear transport \cite{Kaufman_1972,Brizard_2022_q}. However, our present approach takes into account that EP induced transport processes may be induced by a quasi-coherent fluctuation spectrum of non-perturbative nature and characterized by ${\cal O}(1)$ Kubo numbers \cite{zonca2015nonlinear,chen2016physics,zonca2014energetic}, invalidating  fundamental assumptions of quasi-linear theory. Meanwhile, non-local processes that question the Ansatz of local diffusive transport are also accounted for in the present analysis. In fact, by solving the PSZS transport equations through a hierarchy of transport models \cite{zonca_2021} ranging from global gyrokinetics \cite{Bottino_2022} to quasilinear theories, we can develop and validate advanced reduced EP transport models capable of capturing the long timescale evolution of burning plasmas, and provide insights into the non-locality of the underlying transport processes \cite{zonca_2021}.

In order to further articulate and discuss the present gyrokinetic theory of transport in phase-space, we also represent the PSZS evolution equation in the magnetic-drift/banana center frame using standard flux coordinates and the relative shift operator accounting for the guiding center magnetic drift motion in the slowly evolving equilibrium.  The corresponding self-consistent modifications to the reference magnetic equilibrium can be obtained applying the Chew Goldberger Low (CGL) description \cite{chew1956boltzmann}  by means of  the macro-/meso- scopic component of the PSZS moments \cite{cary2009hamiltonian}. The renormalized nonlinear equilibrium evolving on the transport time scale due to self-consistent  fluctuations, sources and collisions is described by the 
zonal state (ZS) \cite{falessi2019, zonca_2021}, consistent with the PSZS transport theory. 
The ZS, thus, consists of the PSZS and its counterpart, i.e., the zonal electromagnetic fields (ZFs), which represent the long-lived component of toroidally symmetric electromagnetic fields. In fact, the ZS does not evolve in the absence of  fluctuations and/or sources and collisions, which is consistent with its definition as a proper nonlinear equilibrium. A more rigorous definition of the ZS is given below in Section \ref{sec:orbitaverage}. The evolution of the ZS discussed in this work expands upon the results in Ref. \cite{falessi2019} and is predominantly due to toroidal symmetry breaking fluctuations. Here, as a further step, we derive an expression for the plasma polarizability that generalizes the expressions derived recently to arbitrary geometry and equilibrium distribution functions, i.e., PSZS \cite{wang__2009,lu_2019,lu_2021,cho_2021}. We also show that transport equation can be cast in the form of a flux surface averaged continuity equation including neoclassical transport in the banana regime as well as sources/sinks. An in-depth discussion of phase-space transport processes due to toroidally symmetry breaking perturbations as well as sources/sinks will be reported in a separate work, where we will also address the possibility of constructing reduced transport models within a unified theoretical framework.  

The article is structured as follows. In Section \ref{sec:orgfa59a06}, we introduce the concept of ZS based on the notion of PSZS and ZFs, which is explored in more detail in Section \ref{sec:orbitaverage}. Next, in Section \ref{sec:renorm}, we demonstrate how PSZS can be interpreted as a renormalization of the reference distribution function in the presence of a finite level of fluctuations. In Section \ref{sec:CGL}, we explore the self-consistent modification of the reference magnetic equilibrium due to the PSZS. Section \ref{sec:org16022e2} focuses on the self-consistent evolution of the ZS, showing how a gyrokinetic transport theory on long time scales can be consistently developed within the present theoretical framework adopting the conservative form of nonlinear gyrokinetic equations and reconnecting to the previous work discussed in Ref. \cite{falessi2019}. Finally, we summarize our findings and discuss future directions in Section \ref{sec:concl-disc}. As further illustrative example of applications of the present theoretical framework, \ref{sec:app} presents to interested readers a detailed discussion of the physics of Geodesic acoustic mode (GAM)/ Energetic particle driven geodesic acoustic mode (EGAM) in general geometry. Although GAM/EGAM do not belong to the ZS as nonlinear equilibrium due to their finite frequency and collisionless damping/drive, their nonlinear dynamics can affect the ZS nonlinear evolution in a peculiar fashion. Thus, they are presented here as particular yet paradigmatic case.

\section{Phase-space zonal structures and the zonal state}
\label{sec:orgfa59a06}
As already mentioned in the Introduction, PSZS are characterized by being ``slowly evolving'' which means that they are not affected by collisionless dissipation, e.g., Landau damping~\cite{zonca2015nonlinear,chen2016physics,zonca2014energetic,falessi2019,zonca_2021}. To satisfy this criterion, PSZS must be calculated by adopting a two-step averaging procedure. More precisely: first, an average along guiding center equilibrium orbits is applied; second, a filter is used, on the axisymmetric particle response, to remove the fast spatiotemporal variations on the characteristic particle orbit length-scale and/or the hydrodynamic time-scale. Consequently, PSZS depend solely on the equilibrium invariants of motion, such as the particle energy (per unit mass) \(\mathcal{E}_{0} = v^{2}/2 \)\footnote{For simplicity of the present analysis, we assume that equilibrium radial electric field, if it exists, corresponds to sufficiently slow $\bm E \times \bm B$ flow that is consistent with the gyrokinetic ordering and, thus, can be incorporated within the perturbed radial electric field. If needed, this assumption could be readily dropped.}, the magnetic moment \(\mu = v_\perp^2/2 B_0 + \ldots \) and the toroidal angular momentum \(P_{\phi}\)~\cite{frieman1982nonlinear}. It is worth noting that any other combination of three invariants of motion can be used, for example involving the `second invariant' $J =  m \oint v_\parallel dl = J (P_\phi, {\cal E}, \mu)$ \cite{zonca2015nonlinear}. In this section, we apply this approach to derive the governing equation for PSZS in conservative form. Furthermore, we derive the equations providing the deviation of the axisymmetric particle response from the PSZS and its dynamic evolution. Then, the notion of PSZS is used to introduce the concept of ZS, which, together with the ZFs defined below in Sec. \ref{sec:orbitaverage}, provides a proper definition of plasma nonlinear equilibrium~\cite{falessi2019,zonca_2021,chen2007nonlinear1} that evolves consistently with the (toroidal) symmetry breaking fluctuation spectrum as well as with sources and collisions. 
\subsection{Orbit averaged particle response: PSZS and zonal state}
\label{sec:orbitaverage}
Since PSZS depend only on the equilibrium constants of motion, their evolution equation can be readily cast using these as coordinates in the phase-space. Proceeding along these lines, we write the phase-space velocity appearing in the gyrokinetic equation~\cite{frieman1982nonlinear,sugama2017modern,brizard2007foundations} as the sum of two contributions, i.e. \(\dot{\bm Z} = \dot{\bm Z}_{0}+ \delta\dot{\bm Z}\), representing, respectively, the integrable particle motion in the reference magnetic field and the remaining particle response that we generically attribute to the effect of fluctuations. This decomposition is general and could be applied to any nearly integrable (Hamiltonian) system. It assumes that the reference equilibrium, defined by the reference magnetic field and by the plasma profiles that are consistent with it and with the PSZS, varies on a sufficiently slow time scale. 
A more rigorous definition of reference or ``equilibrium'' magnetic field, is given in Sec. \ref{sec:CGL}.
The self-consistency of this description and approach can be rigorously checked {\sl a posteriori}. Consequently, the gyrokinetic equation in conservative form reads:
\begin{equation}
\label{eq:6}
\frac{\partial}{\partial t}(D F) + \frac{\partial}{\partial \mathbf{Z}} \cdot\left(D \bm \dot{\bm Z}_{0} F\right) + \frac{\partial}{\partial \mathbf{Z}} \cdot\left(D \delta\bm \dot{\bm Z} F\right) = 0 \, ,
\end{equation}
where $D$ is the velocity space Jacobian and $F$ the gyro-center distribution function~\cite{sugama2017modern,brizard2007foundations}. Here, for the sake of simplicity, we have temporarily suppressed
collisions and source terms. We introduce  \((\theta,\zeta,P_{\phi},\mathcal{E}_{0},\mu)\)
as phase space coordinates\footnote{For circulating particles, the sign of particle motion parallel or anti-parallel to the equilibrium magnetic field must be specified as well, and will be implicitly assumed.} where $\theta$ and $\zeta$ are, respectively, poloidal and toroidal magnetic flux coordinates. Our focus now turns to the zonal distribution function that is the toroidally symmetric part of $F$. This can be obtained by extracting the $n=0$ component of its Fourier expansion, where $n$ is the toroidal mode number. For symmetry reasons, this is the obvious starting point for the definition of an equilibrium distribution function in axisymmetric Tokamak plasmas. Without loss of generality, we assume an axisymmetric equilibrium magnetic field, i.e., $\boldsymbol{B}_0=\hat{F} \boldsymbol{\nabla} \phi+\boldsymbol{\nabla} \phi \times \nabla \psi$ where $\hat{F}=R B_\phi$ and $\phi$ is the toroidal angle, which is connected to $\zeta$ as $\zeta = \phi - \nu(\psi, \theta)$, with $\nu (\psi, \theta)$ chosen such that magnetic flux coordinates are characterized by straight magnetic field lines. We now note that, in the zonal component of Eq. (\ref{eq:6}), we can re-write the term describing the equilibrium motion as: 
\begin{equation}
\label{eq:7}
\frac{\partial}{\partial\bm Z} \cdot \left(D \bm \dot{ \bm Z}_{0} F\right)_{z} = \nabla \cdot \left(D \bm \dot{\bm X}_{0} F\right)_{z} = \frac{1}{{\cal J}_{P_{\phi}}} \frac{\partial}{\partial \theta}(D {\cal J}_{P_{\phi}}F \bm \dot{ \bm X}_{0}\cdot \bm \nabla \theta)_{z} \, ,
\end{equation}
where ${\cal J}_{P_{\phi}} = {\cal J} (\partial P_{\phi}/\partial \psi)^{-1}$, \({\cal J} = (\bm \nabla \zeta \cdot \bm \nabla \psi \times \bm \nabla \theta)^{-1}\) is the Jacobian in flux coordinates, $\psi$ is the poloidal magnetic flux, and the toroidal symmetry of the reference state has been used along with  and the conservation of $P_{\phi}$ and energy characterizing particle motion in the equilibrium magnetic field, i.e., respectively \(\dot{ \bm X}_{0}\cdot \bm\nabla P_{\phi} =0\) and \(\mathcal{\dot{E}}_{0}=0\). We can now orbit average the zonal component of Eq. (\ref{eq:6}) in the reference equilibrium (slowly evolving in time); that is, an averaging operator along \(\theta\) on the gyrokinetic equation while using  \({\cal J}_{P_{\phi}}\)  as weight annihilating, as expected, the term described in Eq. (\ref{eq:7}). Assuming that the reference magnetic equilibrium is slowly evolving; e.g., on the resistive current diffusion time, we finally obtain:
\begin{equation}
\label{eq:2}
\partial_{t} \oint d \theta {\cal J}_{P_\phi} D F_{z} + \oint d \theta {\cal J}_{P_{\phi}} \frac{\partial}{\partial \bm Z} \cdot (D \delta \dot{ \bm Z} F)_{z} = 0 \, .
\end{equation}
Recalling the governing equation  in the absence of fluctuations \cite{cary1983noncanonical,cary2009hamiltonian}; i.e., $\dot{\psi} = -v_{\|} \partial_{\theta}  \bar \psi / ({\cal J} B_{\|}^{*})$, where \(\bar \psi =\psi-R B_\phi v_{ \|} / \Omega  \simeq-(c / e) P_{\phi}\)\footnote{Here, by the $\simeq$ notation, we mean that $\bar \psi$ contains the leading order expression of $P_\phi$. This is not a limitation of the present approach, which can be carried out at the desired order of accuracy consistent with the adopted gyrokinetic formulation \cite{sugama2017modern,brizard2007foundations}.}, $\dot \theta = v_\parallel \partial_\psi \bar \psi/({\cal J} B_\parallel^*)$ and $D = B_\parallel^*/|v_\parallel|$, with $B_{\|}^{*} \equiv \mathbf{B}^{*} \cdot \mathbf{b}$, $\bm b \equiv \bm B_0/B_0$, $\mathbf{B}^{*} \equiv \nabla \times \mathbf{A}^{*}$, $(e/c) \mathbf{A}^{*} \equiv (e/c) \mathbf{A}_{0}+m\left(v_{\|} \mathbf{b}\right)$, $\mathbf{B}_{0}\equiv \nabla \times \mathbf{A}_{0}$, we can recognize that the averaging used to derive Eq. (\ref{eq:2}) is indeed a time averaging along the integrable particle orbit; denoted as
\begin{equation}
\orbave {(\ldots)} = \tau_b^{-1} \oint d \theta (...)/\dot\theta \; , \label{eq:orbave}
\end{equation}
with \(\tau_b = \oint d \theta/\dot\theta\). Thus, we obtain the following (equilibrium) orbit averaged kinetic equation:
\begin{equation}
\label{eq:8}
\frac{\partial}{\partial t}\orbave{F_{z}}+\frac{1}{\tau_{b}}\left[\frac{\partial}{\partial P_{\phi}}\orbave{\left(\tau_{b} \delta \dot{P}_{\phi} F\right)_{z}}+\frac{\partial}{\partial \mathcal{E}}\orbave{\left(\tau_{b} \delta \dot{\mathcal{E}} F\right)_{z}}\right]= \orbave{ \left(\sum_{s} C_s^{g}\left[F_, F_{s}\right]+\mathcal{S}\right)_{z}}  \, ,
\end{equation}
where \(\delta \dot{P}_{\phi} = \delta  \dot{\mathbf{X}} \cdot \bm \nabla P_{\phi}\), $\delta \dot{\mathcal{E}}$ is defined in Eq.(\ref{eq:3}), and we have restored collisions and source terms on the RHS. The expressions of gyrokinetic collision operators of the considered test particles with the field particle species $s$, as denoted by the subscript in $C_s^g$, are given in Refs. 
\cite{sugama2017modern,brizard2007foundations}.
~Meanwhile, for the sake of notation simplicity, the summation over different field particle species in the collisions term will be omitted from now on. Denoting the spatial-temporal slowly evolving component of $\orbave{F_z}$, i.e. PSZS, as $\orbave{F_0} \equiv [\orbave{F_z}]_{S}$, the relevant evolution equation is obtained by additionally extracting the macro-/meso- scopic component of Eq. (\ref{eq:8}); i.e.:
\begin{equation}
\label{eq:120}
\frac{\partial}{\partial t} \orbave{F_{0}}+\frac{1}{\tau_{b}}\left[\frac{\partial}{\partial P_{\phi}} \orbave{\left(\tau_{b} \delta \dot{P}_{\phi} \delta F\right)_{z}}+\frac{\partial}{\partial \mathcal{E}} \orbave{\left(\tau_{b} \delta \dot{\mathcal{E}} \delta F\right)_{z}}\right]_{S}= \orbave{C^{g}}_S + 
\orbave{S}_S \, ,
\end{equation}
where \([\ldots]_{S}\) denotes an appropriate (ad hoc) spatio-temporal averaging procedure to be illustrated; and where we have postulated a bi-linear collision term such that:
\begin{equation}
\label{eq:5}
\orbave{C^{g}}_S=\orbave{C^{g} \left[\orbave{F_{ 0}}, \orbave{F_{ 0}}\right]}+ \left[ \orbave{C^{g} \left[\orbave{F_{ 0}}, \delta F \right]}+ \orbave{C^{g}
\left[\delta F ,\orbave{F_{ 0}} \right]}+\orbave{C^{g} \left[\delta F,  \delta F\right]}\right ]_{S}.
\end{equation}
In the previous expression we have introduced the following decomposition at each instant of time:
\begin{equation}
F = \orbave{F_{ 0}} + \delta F \, .  
\end{equation}
The aforementioned spatio-temporal averaging over the micro-scales is what allows us to separate $\orbave{F_{ 0}}$ from $\orbave{F_{z}}$, given by Eq. (\ref{eq:8}). It is arbitrary, 
to some extent, and can be specified for convenience according to 
the problem of interest. Our choice, here, is to write Eq. (\ref{eq:120}) as definition of $\orbave{F_{ 0}}$, and based on Eq. (\ref{eq:8}), include all residual response in
$\orbave{\delta F_{z}} \equiv \orbave{F_{z}} - \orbave{F_{0}}$ consistent with Eqs. (\ref{eq:28}) and (\ref{eq:dFzave}) below. This point will be further discussed in Sec. \ref{sec:renorm}.
 The approach proposed in the present analysis could be considered a ``full-F'' description \cite{chang2004,zhixin_2023} of the nonlinear evolving equilibrium, and a ``delta-F'' approach \cite{Lee_1981} for the (toroidal) symmetry breaking perturbations (cf. \cite{garbet2010gyrokinetic} for a general review for gyrokinetic simulations of turbulent transport).

Here, it is also worthwhile to briefly remark that the ratio between the third and the second terms of LHS in Eq. (\ref{eq:8}) can be estimated as \(\delta \, \mathcal{E}/\Delta \mathcal{E}\) with \(\delta P_\phi/\Delta P_\phi \sim {\cal O}(1)\), 
where \(\Delta \mathcal{E}\) and \(\Delta P_\phi\) are, respectively, PSZS characteristic scales in energy 
and toroidal angular momentum space; and \(\delta \mathcal{E}\) and \(\delta P_\phi\) are the corresponding nonlinear distortions due to the considered fluctuation spectrum. Using the typical nonlinear gyrokinetic ordering, this is consistent with the relatively small effect of the so-called parallel nonlinearity on fluctuation induced phase-space transport. This is not longer the case for Eq. (\ref{eq:120}), where the two terms are generally of the same order. However, once the effect of the third term is integrated over in energy space, the corresponding overall effect can, again, be shown to be negligible at the leading order. Consistently, in Ref.~\cite{falessi2019} a PSZS transport theory has been formulated omitting the parallel non-linearity term and adopting the classical Frieman-Chen formulation of the nonlinear gyrokinetic equation, which is appropriate up to leading order in the multi-spatiotemporal scale asymptotic expansion~\cite{frieman1982nonlinear}. In the present work, the parallel nonlinearity is retained; consistent with the conservative formulation of nonlinear gyrokinetics~\cite{sugama2017modern,brizard2007foundations}.

\begin{figure}
\centering  
\includegraphics[width=.7\linewidth]{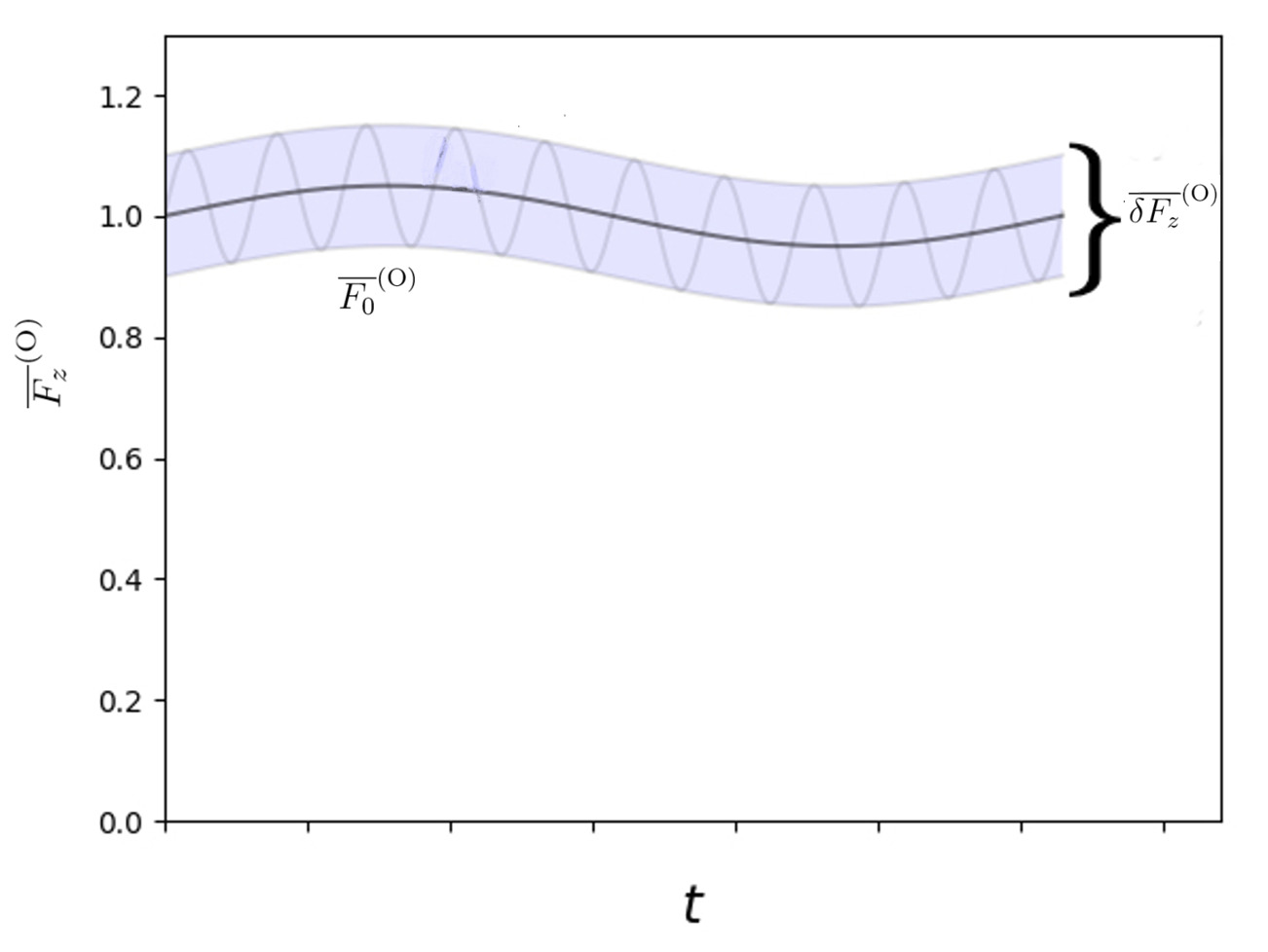}
\caption{Schematic picture describing the equilibrium orbit averaged distribution function $\orbave{F_{z}}$. The solid line represents the slowly varying component of the orbit averaged distribution function, while the oscillation around it corresponds to $\orbave{\delta F_{z}}$.}    
\label{org9968505}
\end{figure}

Having introduced the concept of PSZS, we can  decompose the whole gyrocenter particle response and, consequently, the zonal component of the gyrokinetic distribution \(F_{z}\), into the sum of different terms accounting for the various relevant physics processes, i.e.:
\begin{equation}
\label{eq:28}
F_{z} = \orbave{F_{z}} +\delta \tilde{F}_{z}=\orbave{F_{0}} + \orbave{\delta F_{z}} + \delta \tilde{F}_{z}.
\end{equation}
In particular, as already stated, the PSZS,  \(\orbave{F_{0}}\), describes the evolution of the macro-/meso-scopic equilibrium. The phase-space transport theory, derived in this work is primarily motivated by the fact that this contribution may increasingly deviate in time from the reference thermodynamic equilibrium due to nonlinear processes; eventually invalidating the usual transport analyses that rely, e.g., on local Maxwellian equilibria. Notable applications are analyses of EP transport in fusion plasmas \cite{zonca2015nonlinear,chen2016physics,zonca2014energetic}, but deviation of (bulk) particle equilibria from local Maxwellian can be also important to explain, e.g., the nonlinear up-shift (the so-called ``Dimits shift'' \cite{dimits_2000}) of the threshold for ion temperature gradient driven turbulence \cite{chen2007nonlinear1}. In the following, we will show that a multipole expansion can be applied to the PSZS fluid moments~\cite{brizard2007foundations,cary2009hamiltonian} yielding an anisotropic CGL pressure tensor~\cite{chew1956boltzmann,kulsrud83} and a self-consistently evolving nonlinear magnetic equilibrium. 
Further to \(\orbave{F_{0}}\), the residual components of \(F_{z}\) either describe the micro-scale spatio-temporal variation of the orbit averaged distribution function or have zero average along equilibrium orbits. More precisely, the residual orbit averaged particle response, \(\orbave{\delta F_{z}}\), characterizes the transition between neighboring nonlinear equilibria, which are all undamped by collisionless dissipation~\cite{falessi2019,chen2007nonlinear1} and slightly deviate from the reference $\orbave{F_0}$ as schematically described in Fig.~\ref{org9968505}.
These neighboring nonlinear equilibria~\cite{chen2007nonlinear1} should be understood as the ensemble of different realizations of the system, thereby 
representing the connection between time average, introduced above in the definition of $\orbave{F_0}$ by means of Eq. (\ref{eq:120}), and ``ensemble average'' 
in a statistical sense~\cite{falessi2019}.
The distribution function \(\orbave{F_{z}} {= \orbave{F_{0}} + \orbave{\delta F_{z}}}\), together with the undamped components of the electromagnetic potentials (ZFs), constitute the ZS, i.e., ``the collisionless undamped (long-lived) nonlinear deviation of the plasma from the reference state as a consequence of fluctuation-induced transport processes, due to emission and reabsorption of (toroidal equilibrium) symmetry-breaking perturbations'' \cite{falessi2019}. 
We note that the symmetry breaking fluctuations (with $n \neq 0$) are not explicitly mentioned as elements of the ZS, 
but they are self-consistently accounted for in its definition,
as they have determined  \(\orbave{F_{z}}\) during the dynamic evolution of the system. We also note that, although PSZS can be obtained separating out the macro-/meso-scopic component of \(\orbave{F_{z}}\) 
consistently with the usual definition of equilibrium and transport, this separation is somehow arbitrary and depends on the specific problem of interest. For example, when describing EPM~\cite{chen1994epm}, \(\orbave{F_{z}} = \orbave{F_{0}} + \orbave{\delta F_{z}}\) is best considered as a whole since in this case
phase-space transport occurs on the same time scale of the nonlinear dynamic evolution of the spectrum of fluctuations~\cite{zonca2015nonlinear,chen2016physics}. Following the previous derivation, we obtain the governing equation for the spatio-temporal micro-scale component of the equilibrium orbit averaged distribution function:
\begin{multline}
\label{eq:dFzave}
\frac{\partial}{\partial t} \orbave{\delta F_z} + \frac{1}{\tau_{b}}\left[\frac{\partial}{\partial P_{\phi}} \orbave{\left(\tau_{b} \delta \dot{P}_{\phi}\orbave{ F_{0}} \right)_{z}}+\frac{\partial}{\partial \mathcal{E}} \orbave{\left(\tau_{b} \delta \dot{\mathcal{E}} \orbave{ F_{0}}\right)_{z}}\right] \\  +\frac{1}{\tau_{b}}\left[\frac{\partial}{\partial P_{\phi}} \orbave{\left(\tau_{b} \delta \dot{P}_{\phi} \delta F\right)_{z}}+\frac{\partial}{\partial \mathcal{E}} \orbave{\left(\tau_{b} \delta \dot{\mathcal{E}} \delta F\right)_{z}}\right]_{F} = [\orbave{C^{g}}_{z} + \orbave{S}]_{F}.
\end{multline}
Note that, here, \([\ldots]_{F}\) denotes the  spatio-temporal micro-scale component of the argument such that \([\ldots] \equiv [\ldots]_{S} + [\ldots]_{F}\); that is, the spatial variation on Larmor radius and finite magnetic drift orbit width length scale, and the temporal variation on the hydrodynamic time scale. 
It should be pointed out that, as opposed to the governing equation for PSZS, there is a formally linear term in the orbit averaged response on the LHS. This term 
may have fast as well as slow spatio-temporal variations and, thus, the subscript $F$ is omitted there. Furthermore, this same term
is also responsible for the high frequency oscillation characterizing the geodesic acoustic mode (GAM)~\cite{winsor1968} and, therefore, cannot be included into the definition of a macroscopic equilibrium consistent with usual transport time scale orderings \cite{falessi2019}. Interested readers can find linear as well as nonlinear GAM/EGAM physics discussed in detail in \ref{sec:app}.

\subsection{Nonlinear equilibrium as renormalized particle response} \label{sec:renorm}
In the previous subsection, we showed that the concept of PSZS is intrinsically related to the integrable ``equilibrium'' guiding center motion and, thus, it is naturally described using \(P_{\phi}\) as phase-space coordinate. In particular, the governing equations for the different components of the zonal distribution function take very compact expressions. 
However, when describing the self-consistent evolution of ZFs, we must adopt  standard flux coordinates  \((\psi,\theta,\zeta)\). We can define the associated change of coordinates between these two representations noting that \(P_\phi =P_{\phi}(\psi, \theta, \mathcal{E}, \mu)= -(e/c) (\psi - \delta \tilde \psi(\psi ,\theta, \mathcal{E}, \mu) )\). Thus, consistently with previous works~\cite{hinton1999dynamics,zonca2015nonlinear,falessi2019}, 
one can introduce a shift operator, formally represented as \(e^{iQ}\), accounting for the guiding center equilibrium magnetic drift motion, which therefore provides the push-forward transformation from gyrocenter to magnetic drift/oscillating-centers. Then, the (equilibrium) orbit average of a scalar function $\hat H (P_\phi, \theta) = H ( \bar \psi  + \delta \tilde \psi , \theta)$ reads:
\begin{equation}
\label{eq:1}
\oint \frac{d \theta}{\dot{\theta}} \hat H (P_\phi, \theta) = \oint \frac{d \theta}{\dot{\theta}} H ( \bar \psi  + \delta \tilde \psi , \theta) = \oint \frac{d \theta}{\dot{\theta}} e^{i Q} H ( \bar \psi , \theta)
\end{equation}
where \(\bar{\psi} = - (c/e) P_\phi\) was defined in Sec. \ref{sec:orbitaverage} above and $\delta \tilde \psi = R B_\phi v_\parallel/\Omega$ at the leading order, with its dependence on $\theta$ and the other phase-space variables left implicit. It follows by direct inspection that, as expected, (equilibrium) orbit averaging  is equivalent to a bounce/transit average combined with the action of the shift operator \(e^{iQ}\). As a further remark, we recall that  bounce/transit averaging is also connected with flux surface averaging of velocity space integrals and, consequently, with the standard representation of plasma (radial) transport equations. For this reason, the PSZS governing equation is particularly relevant for describing plasma transport and allows recovering well known results~\cite{abel2013multiscale}, and further generalizing them~\cite{falessi2019}. In order to see the equivalence between orbit averaging and ``shifted bounce 
averaging'' more clearly, let us define \(\overline {(\ldots)} = \tau_b^{-1} \oint d \theta (...)/\dot\theta\), where, now, 
the closed poloidal orbit 
integral follows the  constant-$\theta$ projection of the actual guiding center orbit on the $\bar \psi$ flux surface. This definition of \(\tau_b\)
with respect to the orbit averaging approach, is unambiguous since it is uniquely defined being $\theta$ a dummy integration variable. Then, 
\begin{equation}
\label{eq:equivalave}
\left. \orbave {(\ldots)} \right|_{P_\phi} = \left. \overline {e^{i Q} (\ldots)} \right|_{\bar \psi} \, , \end{equation}
where, for further clarity, we have explicitly denoted by the additional subscripts the reference value of $P_\phi$ on the LHS, and of $\bar \psi$ on the RHS. 
Rephrasing this concept, Eq. (\ref{eq:equivalave}) states that orbit average for given $P_\phi$ (implicitly assuming given ${\cal E}$ and $\mu$) corresponds
to a proper shifted bounce averaging on the flux surface labeled by $\bar \psi$.

Due to the equivalence between orbit and bounce/transit averaging, the governing equation for PSZS introduced in the previous subsection are consistent with those obtained in Refs. \cite{zonca2015nonlinear,falessi2019}, with the further inclusion of the effects of the so called parallel nonlinearity. In what follows, we re-derive the governing equations for the different components of $F_{z}$ using \((\psi, \theta, \zeta,\mathcal{E}, \mu)\)
as phase-space coordinates for two main reasons: first, in order to establish a close contact with the representation that is most conveniently adopted for writing equations describing mode structures, either ZFs or symmetry breaking perturbations; and second, for demonstrating that the PSZS definition adopted in this and earlier works~\cite{zonca2015nonlinear,chen2016physics,zonca2014energetic,falessi2019} corresponds to a renormalization of equilibrium particle response.

As a first step towards the second goal, we note that the decomposition of Eq.~\eqref{eq:28} can be written by introducing the drift/banana center pull-back operator \(e^{-i Q_{z}}\) where \(Q_{z}=R B_\phi \left(v_{\|} / \Omega\right) k_{z} /(d \psi / d r)\)~\cite{falessi2019,zonca_2021,chen2007nonlinear1,hinton1999dynamics}. This is the explicit expression for the shift operator introduced previously and the subscript \(z\) to the radial wave number $k_z \equiv - i \partial_r$ reminds that it is acting on toroidally symmetric response. As noted above,  \(Q_{z}\) is the leading order expression of \(Q\); and more accurate expressions for  \(Q_{z}\) could be given by means of corresponding more accurate expressions of $P_\phi$, consistent with the adopted gyrokinetic description \cite{sugama2017modern,brizard2007foundations}. In  particular, \(F_{z}\) can be written as:
\begin{equation}
\label{eq:36}
F_{z} \equiv \overline{F}_{0}+e^{-i Q_{z}}\left(\overline{\delta F_{B z}}\Big|_{F}+\delta \tilde{F}_{B z}\right)=\overline{F}_{0}+e^{-i Q_{z}}\left(\left.\overline{e^{i Q_{z}} \delta F_{z}}\right|_{F}+\delta \tilde{F}_{B z}\right)
\end{equation}
where the function \(\delta F_{Bz}\) is the drift/banana center particle response, the bar stands for  bounce/transit averaging, the tilde denotes 
the vanishing bounce/transit average response and \(\bar{F}_{0}\) is, from now on, a short notation
for $\orbave{F}_0$ and, thus, describes the PSZS component. 
Note the one-to-one correspondence of Eq. (\ref{eq:36}) and Eq. (\ref{eq:28}), which also illuminates the notation.
Following the previous subsection, we now proceed in deriving the governing equations for the different terms of this decomposition. In particular, 
we recall the gyrokinetic expression for the time variation of the energy per unit mass, i.e.:
\begin{equation}
\label{eq:3}
\delta \dot{\mathcal{E}}=-\frac{e}{m} \dot{\bm X}_{0} \cdot \nabla\left\langle\delta L_{g}\right\rangle_{z} \, ,
\end{equation}
where angular brackets denote gyro-phase averaging, $\delta L = \delta \phi - \bm v \cdot \delta \bm A /c$, $\delta \phi$ is the scalar potential, $\delta \bm A$ is the vector potential  with $\delta L_g = e^{\bm \rho \cdot \bm \nabla} \delta L (\bm X) = \delta L (\bm X + \bm \rho)$, $\bm \rho = \bm b_0 \times \bm v/\Omega$ and $\Omega = e B_0/(mc)$. Note that, $\left\langle \delta L_g \right\rangle_z = J_0(\lambda) (\delta \phi_z - v_\parallel \delta A_{\parallel z} ) + (2/\lambda) (m/e) \mu J_1(\lambda) \delta B_{\parallel z}$, $\lambda^2 = 2 (\mu B_0/\Omega^2) k_\perp^2$ and $J_{0,1}$ are Bessel functions. We also recall the conservation of the toroidal component of the canonical angolar momentum in the presence of toroidally symmetric perturbations:
\begin{equation}
\label{eq:4}
\delta \dot{\theta} \frac{\partial P_{\phi}}{\partial \theta}+\delta \dot{\psi} \frac{\partial P_{\phi}}{\partial \psi} + \delta \dot{\mathcal{E}} \frac{\partial P_{\phi}}{\partial \mathcal{E}} = - \frac{e}{c} \left (\partial_{t} + \dot{\bm X}_{0} \cdot \nabla\right) \frac{R B_\phi \langle \delta A_{\parallel g}\rangle_{z}}{B_{0}}\;.
\end{equation}
Thus, we can re-write the toroidally symmetric component of Eq.~\eqref{eq:6} as follows:

\begin{align}
\label{eq:38b}
  D(&\partial_{t} + \dot{\bm X}_{0}\cdot \bm \nabla)  \left(F_{z}- \frac{e}{m} \langle \delta L_{g}\rangle_{z} \frac{\partial \bar{F}_{0}}{\partial \mathcal{E}} \Big |_{\bar{\psi}} + \frac{R B_\phi}{B_{0}} \frac{\partial \bar{F}_{0}}{\partial \bar{\psi}} \langle \delta A_{\parallel g}\rangle_{z}\right) + \notag \\
  &
  { - D \frac{R B_\phi}{B_{0}} \langle \delta A_{\parallel g}\rangle_{z} \frac{\partial}{\partial \bar{\psi}} \partial_t \bar{F}_{0}}+  D \frac{e}{m} { \frac{\partial}{\partial t} \left( \left.\frac{\partial \bar{F}_{0}}{\partial \mathcal{E}}\right|_{\bar{\psi}} \langle \delta L_{g}\rangle_{z} 
    \right) }+ \notag \\
  & +\frac{1}{{\cal J}} \frac{\partial}{\partial \theta}({\cal J} D \delta \dot{\theta} \delta F) +  \frac{1}{{\cal J}} \frac{\partial}{\partial \psi}({\cal J} D \delta \dot{\psi} \delta F) +  \frac{\partial}{\partial \mathcal{E}}( D \delta \dot{\mathcal{E}} \delta F)  =D (C^{g} + \mathcal{S}).
\end{align}

This equation, consistently with Ref.~\cite{zonca_2021,frieman1982nonlinear}, suggests to introduce the following definition:
\begin{equation}
\label{eq:31}
G_{z} \equiv F_{z}-\left.\frac{e}{m}\left\langle\delta L_{g}\right\rangle_{z} \frac{\partial \bar{F}_{0}}{\partial \mathcal{E}}\right|_{\bar{\psi}} +\frac{R B_\phi}{B_{0}}\left\langle\delta A_{\| g}\right\rangle_{z} \frac{\partial \bar{F}_{0}}{\partial \bar{\psi}} 
\end{equation}
where, as radial coordinate, we are using \(\bar{\psi} \equiv-(c / e) P_{\phi}\) introduced earlier. From the definition above, the role of $\bar{F}_{0}$ as renormalized reference distribution function taking into account nonlinear plasma behaviors (self-interactions) consistently with the theoretical framework introduced in \cite{zonca2015nonlinear,falessi2019} is made clear. In fact, consistently with Eqs. (\ref{eq:28}) and Eq. (\ref{eq:36}),
no distinction is made in Eqs. (\ref{eq:38b}) and (\ref{eq:31}) between the $\orbave{\delta F_{z}} = \overline{e^{iQ_z}\delta F_{z}}$ contribution that should be kept distinct from $\bar{F}_{0}$ and the one that can be reabsorbed in it. Thus, the distinction
can be made for convenience of identification of a reference magnetic equilibrium involving macro- and meso-scale kinetic profiles only (cf. next subsection); but, the 
physics analysis of phase-space structures that are undamped by linear collisionless processes is ``full-$F$'' by construction.
We may also note that, consistently with Eqs. (\ref{eq:28}) and Eq. (\ref{eq:36}), the reference state appearing in Eq. (\ref{eq:31}) 
generally includes a  spatio-temporal micro-scale contribution.  However, consistently with the gyrokinetic ordering~\cite{frieman1982nonlinear,brizard2007foundations}, this term can be neglected at the relevant leading order. We can now write \(G_{z}\) in terms of the drift/banana shift operator, i.e. \(G_{z} = e^{-iQ_{z}}G_{Bz}\), substitute this expression in Eq.~\eqref{eq:38b} and apply \(e^{iQ_{z}}\) on both sides. We find:
\begin{equation}
\label{eq:9}
e^{iQ_{z}} D \dot{\bm X}_{0}\cdot \bm \nabla G_{z}
 = e^{iQ_z} \frac{v_\parallel}{{\cal J} |v_\parallel|}\left [ 1 - \frac{\partial}{\partial \psi} \frac{R B_\phi v_{\parallel}}{\Omega}\right] e^{-iQ_z} \frac{\partial G_{Bz}}{\partial \theta}  =
 \frac{v_\parallel}{{\cal J_{\bar \psi}}|v_\parallel|} \frac{\partial}{\partial \theta} G_{Bz} \, ,
\end{equation}
where ${\cal J}_{\bar \psi} =- (e/c){\cal J_{P_\phi}}$ is computed at the actual gyrocenter particle position; and we have noted $i Q_{z} = (\psi - \bar \psi) \partial_{\psi}$. Considering the effect of the shift operator on \(D\), we obtain the following kinetic equation:
\begin{multline}
{{\cal J}_{\bar \psi} D}\partial_{t}( G_{Bz})  + \frac{v_\parallel}{|v_\parallel|}  \partial_{\theta}G_{Bz} = \\ e^{iQ_{z}}\Big [- \frac{e}{m} 
{\cal J}_{\bar \psi} D \frac{\partial}{\partial t} \left(\langle \delta L_{g}\rangle_{z} \frac{\partial \bar F_0}{\partial \mathcal{E}}\Big |_{\bar{\psi}}   
\right) +  {\cal J}_{\bar \psi} D \frac{R B_\phi}{B_0} \langle \delta A_{\parallel g}\rangle_{z} \frac{\partial}{\partial \bar \psi} \partial_t \bar F_0 \Big]\\ -  \frac{\partial}{\partial \theta}(e^{i Q_{z}}{\cal J}_{\bar \psi} D \delta \dot{\theta} \delta F) - \frac{\partial}{\partial \psi} ( e^{iQ_{z}}
{\cal J}_{\bar \psi} D \delta \dot{ \bar \psi} \delta F) -  \frac{\partial}{\partial \mathcal{E}}(e^{i Q_{z}}{\cal J}_{\bar \psi} D \delta \dot{\mathcal{E}} \delta F) + 
e^{i Q_{z}}{\cal J}_{\bar \psi} D (C_{g} + \mathcal{S}).
\end{multline}
In fact, it can be shown that the shift operator commutes with the partial derivatives 
of nonlinear terms in Eq. (\ref{eq:38b}) taking into account formal simplifications among commutators. Note that $\delta \dot{\bar \psi}$ is typically dominated by $\delta \dot{\psi}$ for $n \neq 0$ symmetry breaking perturbations. The nonlinearities caused by ZFs, meanwhile, need a special attention, since $\delta \dot \psi_z$ vanishes at the leading order. We will come back to this technical but important point while describing some applications of this theory in \ref{sec:app}. Integrating over \(\theta\) on a closed trajectory, the \(\theta\) derivatives can be annihilated. Recalling the bounce average definition introduced previously, the following expression is finally obtained:
\begin{equation}
\label{eq:16a}  
\begin{aligned}
\partial_{t} \overline{G_{B z}}=&-\overline{e^{i Q_{z}} \frac{e}{m} \partial_{t}\left[\left.\left\langle\delta L_{g}\right\rangle_{z} \frac{\partial \bar{F}_{0}}{\partial \mathcal{E}}\right|_{\bar{\psi}} \right]}+\overline{e^{i Q_{z}} \frac{R B_\phi}{B_{0}}\left\langle\delta A_{\| g}\right\rangle_{z} \frac{\partial}{\partial \bar{\psi}} \partial_{t} \bar{F}_{0}}+\left.\overline{e^{i Q_{z}}\left[C_{g}+\mathcal{S}\right]}\right|_{z} \\
&-\frac{1}{\tau_{b}} \frac{\partial}{\partial \psi}\left[\tau_{b} \overline{e^{i Q_{z}} \delta \dot{\bar \psi} \delta F}\right]_{z}-\frac{1}{\tau_{b}} \frac{\partial}{\partial \mathcal{E}}\left[\tau_{b} \overline{e^{i Q_{z}} \delta \dot{\mathcal{E}} \delta F}\right]_{z}
\end{aligned}
\end{equation}
This is a generalization of Eq. (25) in Ref.~\cite{falessi2019}  written in conservative form and retaining the role of parallel nonlinearity, collisions and source terms. Consequently, recalling the relationship between bounce/transit and flux surface averaging, from this expression it is possible to derive all the usual flux surface averaged transport equations~\cite{falessi2019}. The governing equation for \(\bar{F}_{0}\) follows directly from Eq.~\eqref{eq:16a} and is consistent with Eq. \eqref{eq:120}:
\begin{multline}
\label{eq:PSZSev1}
\partial_t \overline{e^{i Q_z} \bar F_0} =  - \left. \overline{e^{i Q_z}\frac{R B_\phi}{B_0} \partial_t \left\langle 
\delta A_{\parallel g}  \right\rangle_z \frac{\partial}{\partial \bar \psi } \bar F_{0}} \right|_S + \left. \overline{e^{i Q_z} \left[ C_g + {\cal S} \right]} \right|_{z S} + 
\\ - \frac{1}{\tau_b} \frac{\partial}{\partial \psi} \left[ \tau_b \overline{e^{i Q_z} 
\delta \dot{\bar  \psi} \delta F} \right]_{z S} - \frac{1}{\tau_b} \frac{\partial}{\partial {\cal E}} \left[ \tau_b \overline{e^{i Q_z} 
\delta \dot {\cal E} \delta F} \right]_{z S} \; .
\end{multline}
Here, it is worthwhile noting that, except for the first (ZFs inductive) term on the RHS, Eq.~\eqref{eq:PSZSev1} shows that PSZS evolution is either caused by nonlinear interactions or by sources/collisions. Physically, the ZFs inductive term is due to the externally or nonlinearly generated perturbation of the magnetic flux function. Thus, the present definition of PSZS, consistent with earlier works~\cite{zonca2015nonlinear,chen2016physics,zonca2014energetic,falessi2019}, describes the renormalized reference distribution function taking into account nonlinear plasma behaviors (self-interactions). The orbit averaged fast spatiotemporal deviation of the plasma response about the PSZS is given by:
\begin{align}
\label{eq:deltagf}        
\partial_t \left. \overline{\delta g_{Bz}} \right|_F & =  - \left. \overline{e^{i Q_z} \frac{e}{m} \partial_t \left[\left\langle \delta L_g \right\rangle_z \left. 
\frac{\partial}{\partial {\cal E}}\right|_{\bar \psi} \bar F_{0} \right]} \right|_F
+ \left. \overline{e^{i Q_z}\frac{R B_\phi}{B_0}  \left\langle 
\delta A_{\parallel g}  \right\rangle_z \frac{\partial}{\partial \bar \psi }\partial_t \bar F_{0}} \right|_F \notag \\
&+ \left. \overline{e^{i Q_z} \left[ C_g + {\cal S} \right]} \right|_{z F} - \frac{1}{\tau_b} \frac{\partial}{\partial \psi} \left[ \tau_b \overline{e^{i Q_z} 
\delta \dot{\bar \psi} \delta F} \right]_{z F} - \frac{1}{\tau_b} \frac{\partial}{\partial {\cal E}} \left[ \tau_b \overline{e^{i Q_z} 
\delta \dot {\cal E} \delta F} \right]_{z F}\; .
\end{align}
where $\delta g_z = e^{-i Q_z} \delta g_{B z} $, consistent with Eq. (\ref{eq:31}), is the nonadiabatic 
particle response that is connected with $\delta F_z$ by:
\begin{equation}
\label{eq:11}
\delta g_z \equiv \delta F_z -\left.\frac{e}{m}\left\langle\delta L_{g}\right\rangle_z \frac{\partial \bar F_0}{\partial \mathcal{E}}\right|_{\bar{\psi}} +\frac{R B_\phi}{B_{0}}\left\langle\delta A_{\| g}\right\rangle_z \frac{\partial \bar F_0}{\partial \bar{\psi}} \; .
\end{equation}
Similarly, after some lengthy but straightforward algebra, one can obtain the governing equation for $\delta \tilde g_{B z} =  \delta g_{B z} - 
\overline{\delta g_{B z}}$ \cite{zonca_2021}. 
Equation \eqref{eq:PSZSev1}, or the equivalent Eq. \eqref{eq:120}, and Eq. \eqref{eq:deltagf} completely describe the ZS, introduced and defined in the previous subsection, once the reference magnetic equilibrium and the evolution equations for the ZFs are given along with the symmetry breaking fluctuation spectrum. This is done in the next subsection.
In particular, the ZS, consisting of neighboring nonlinear equilibria \cite{chen2007nonlinear1} which, can be thought of as ensemble of different
realizations of the system \cite{falessi2019}, can be written as \cite{zonca_2021}
\begin{equation}
 F_{0*} \equiv \bar F_0 + e^{-i Q_z} \left. \overline{e^{iQ_z} \delta F_z } \right|_F . \label{eq:F0*}
\end{equation}

\subsection{Chew Goldberger Low reference equilibrium and zonal fields}
\label{sec:CGL}
The motivation for analyzing phase-space features of transport processes in low collisionality burning plasmas is that PSZS could significantly deviate from a given model plasma equilibrium, e.g. Maxwellian, over long time scales and, thus, the usual transport description as evolution of macroscopic radial profiles may become inadequate~\cite{falessi2019}. Consistently with these motivations, it is necessary to describe the modification of the reference magnetic equilibrium self-consistently with PSZS. Following Refs.~\cite{chen2016physics,cary2009hamiltonian,brizard2007foundations}, we recall that the transformation from the gyrocenter to the particle distribution function can be cast as:
\begin{equation}
\label{eq:23}
\begin{aligned}
f=& e^{-\rho \cdot \nabla} \left[ {F}-\frac{e}{m} \left(\frac{\partial {\overline{F_{0}}}}{\partial \mathcal{E}}+\frac{1}{B_{0}} \frac{\partial {\overline{F_{0}}}}{\partial \mu}\right) \left\langle\delta L_{g}\right\rangle
\right] +\frac{e}{m}\left[  \frac{\partial {\overline{F_{0}}}}{\partial \mathcal{E}} \delta \phi + 
  \frac{1}{B_{0}} \frac{\partial {\overline{F_{0}}}}{\partial \mu} \delta L
\right].
\end{aligned}
\end{equation}
Using this expression, we can write every fluid moment  in terms of its so called push-forward representation~\cite{brizard2007foundations}. In the following, differently from the usual approach, we will describe the guiding center transformation using the Dirac delta formalism instead of the \(e^{- \bm \rho \cdot \bm \nabla}\) operator, extending to phase-space the velocity space integrals on the particle distribution function, expressed as in Eq. \eqref{eq:23}. As a simple example, the toroidally symmetric plasma current density \(\mathbf{J}_{z}\) reads:
\begin{eqnarray}
\label{eq:dJz}
 & & \hspace*{-5em} \mathbf{J}_{z}(\mathbf{r}) = e \int d \mathcal{E} d\mu d\alpha d^{3}  \mathbf{X}\, D\left ( T^{-1}_{gc} \mathbf{v} \right) \delta(\mathbf{X} + \bm \rho- 
\mathbf{r}) \Big[ \bar F_{0} + \delta F_{z} - \frac{e}{m}  \frac{\partial \bar F_{0}}{\partial \mathcal{E}} \langle \delta L_{g}\rangle_{z} +
\nonumber
 \\  & & \hspace*{-2em} -  \frac{e}{m} \frac{1}{B_0} \frac{\partial \bar F_{0}}{\partial \mu} \langle \delta L_{g}\rangle_{z} \Big] +\frac{e^{2}}{m} \int d \mathcal{E} d\mu d\alpha d^{3}  \mathbf{X}\,\mathbf{v} D \Big[  \frac{\partial \bar F_{0} }{\partial \mathcal{E}}  \delta \phi_{z} + \frac{1}{B_{0}}\frac{\partial \bar F_{0}}{\partial \mu}  \delta L_{z} \Big] \; ,
\label{eq:dJz} \end{eqnarray}
where $\alpha$ is the gyrophase, $T^{-1}_{gc} \mathbf{v}$ represents the guiding-center transformation of the velocity $\mathbf{v}$ and the argument of the delta function accounts for the relation between the particle position $\mathbf{r}$ and the guiding center position $\mathbf{X}$~\cite{cary2009hamiltonian}. The pressure tensor can be derived analogously. In the present approach, ZFs are considered explicitly as a distortion of the nonlinear equilibrium, that is of the zonal state. Thus, the reference magnetic equilibrium must be computed assuming only the PZSZ as describing the reference state; i.e., only the $\propto \bar F_0$ term in the push forward representation of the fluid moments such as Eq. (\ref{eq:dJz}). Due to the macro-/meso-scopic nature of PSZS, and applying the usual multipole expansion in the push-forward representation of the fluid moments~\cite{brizard2007foundations,cary2009hamiltonian}, one can obtain a CGL pressure tensor and a toroidally symmetric current satisfying the following force balance equation:
\begin{equation}
\label{eq:35}
\sigma \frac{\mathbf{J}_{z} \times \mathbf{B}_{0}}{c}=\nabla P_{\|}+(\sigma-1) \nabla\left(\frac{B_{0}^{2}}{8 \pi}\right)+\frac{B_{0}^{2}}{4 \pi} \nabla_{\perp} \sigma
\end{equation}
where $\perp$ and $\parallel$ denote the components perpendicular and parallel to $\bm B_0$ and
\begin{equation}
\label{eq:38}
\sigma=1+\frac{4 \pi}{B_{0}^{2}}\left(P_{\perp}-P_{\|}\right).
\end{equation}
It is well-known that, assuming $\bm B_0 = \hat F \nabla \phi + \nabla \phi \times \nabla \psi$, the radial component of this expression reads:
\begin{equation}
\label{eq:39}
\Delta^{*} \psi+\boldsymbol{\nabla} \ln \sigma \cdot \boldsymbol{\nabla} \psi=-\frac{4 \pi R^{2}}{\sigma} \frac{\partial P_\parallel}{\partial \psi} -\frac{1}{\sigma^{2}} \frac{\partial G}{\partial \psi}
\end{equation}
where $\Delta^*$ is the usual Grad-Shafranov operator and $G(\psi)=(\sigma \hat F)^2/2$ is a flux function. Meanwhile, pressure components and $\hat F(\psi, B_0)$ function are connected by 
the parallel 
\begin{equation}
\frac{\partial P_\parallel}{\partial B_0} = \frac{P_\parallel - P_\perp}{B_0}
\label{eq:38par}
\end{equation}
and bi-normal 
\begin{equation}
\frac{\partial P_\perp}{\partial B_0} = \frac{P_\perp-P_\parallel}{B_0} - \sigma \frac{B_0^2}{4\pi} \frac{\partial \ln \hat F}{\partial B_0}
\label{eq:38binorm}
\end{equation}
components of Eq. (\ref{eq:35}). The resulting solution of Eqs. (\ref{eq:35}) to (\ref{eq:38binorm}) defines the magnetic equilibrium that is consistent with the presence of PSZS, which $\bm J_z$ as well as $P_\perp$ and $P_\parallel$ have been computed from. More precisely, $P_{\perp}$ and $P_{\parallel}$ can be calculated integrating the PSZS distribution function and, then, $\hat F(\psi,B_0)$ is obtained from the expression of the poloidal plasma current, which is also computed from $\bar F_0$, and from Eq.(\ref{eq:38binorm}). Finally a standard Grad Shafranov problem must be solved . This produces, as expected \cite{cary2009hamiltonian,brizard2007foundations,kulsrud1980}, an anisotropic MHD equilibrium. It is worth noting that this result holds at the leading order in the multipole expansion. At higher order, we could compute the macro-/meso-scopic deviations from the CGL pressure tensor, which is expected to become relevant when the multipole expansion does not hold; \exgra, when steep gradient regions are encountered, viz., near the last closed magnetic surface. More generally, whenever the length scale of the gradients becomes comparable with the characteristic length of particle orbits, the proposed separation of scales to isolate the PSZS is no longer applicable and a ``full-$F$'' approach is mandatory~\cite{murugappan2022gyrokinetic}. 


The neighboring nonlinear equilibria \cite{zonca_2021,chen2007nonlinear1,falessi2019}; that is, the micro spatiotemporal deviation from the reference state given by PSZS and the just constructed anisotropic (CGL) reference magnetic equilibrium, can also be self-consistently determined along with the ZFs; \idest, $\delta \phi_z$, $\delta A_{\parallel z}$ and $\delta B_{\parallel z}$ are obtained by means of quasineutrality and Amp\`ere equations following the well known theoretical framework described in~\cite{chen2016physics}:
\begin{align}
    &    \sum_s\left\langle\frac{e^{2}}{m} \frac{\partial \bar{F}_{0s}}{\partial \mathcal{E}}\right\rangle_{v} \hspace{-0.3em}\delta \phi_{z}\hspace{-0.3em}+\nabla \cdot \sum_s\left\langle\frac{e^{2}}{m} \frac{2 \mu}{\Omega^{2}} \frac{\partial \bar{F}_{0s}}{\partial \mu}\left(\frac{J_{0}^{2}-1}{\lambda^{2}}\right)\right\rangle_{v} \nabla_{\perp} \delta \phi_{z} \nonumber \\
    & \hspace*{2em} + \sum_s\left\langle e J_{0}(\lambda) \delta g_{z}\right\rangle_{v} +  \sum_s\left\langle e J_{0}(\lambda) \bar F_{0s} \right\rangle_{v} =0 \; , \label{zfs1}\\
   & 
     \frac{\partial}{\partial t} \delta A_{ \| z}= - \left[\frac{1}{B_{0}} \boldsymbol{b}_{0} \times \boldsymbol{\nabla} \delta A_{ \|} \cdot \boldsymbol{\nabla} \left( \nabla_\parallel^{-1} \partial_t \delta A_\parallel \right) \right]_{z} \; , \label{zfs2}\\
  & 
\nabla_{\perp} \delta B_{\| z} = \bm \kappa_{0} \delta B_{\parallel z} + \nabla_{\parallel} \delta \mathbf{B}_{\perp z} + \nabla \mathbf{b}_{0}\cdot \delta \mathbf{B}_{\perp z} + \frac{4 \pi}{c} \delta \mathbf{J}_{\perp z}\label{zfs3}.
\end{align}
Here, $\sum_{s}$ denotes summation on all particle species, $\left\langle ... \right\rangle_v$ stands for velocity space integration, $\nabla_\parallel^{-1}$ is the inverse operator of $\nabla_\parallel$, $\bm \kappa_0 \equiv \bm b_0 \cdot \bm \nabla \bm b_0$ 
is the magnetic field curvature, $\delta \mathbf{J}_{\perp z}$ is readily obtained from Eq. (\ref{eq:dJz}), 
$\delta \bm B_{\perp z}$ and $\delta B_{\parallel z}$ are expressed in terms of the fluctuating vector potential 
as in Ref. ~\cite{chen2016physics}, and the Coulomb gauge $\bm  \nabla \cdot \delta \bm A = 0$ is assumed. Thus, 
\begin{eqnarray}
\delta \bm B_{\perp z} & = & \bm \nabla_\perp \delta A_{\parallel z} \times \bm b_0 + \bm b_0 \times \bm \kappa_0 \delta A_{\parallel z} \nonumber \\
& & + \bm b_0 \times \nabla_\parallel \delta \bm A_{\perp z} + (\bm b_0 \times \bm \nabla \bm b_0 ) \cdot \delta \bm A_{\perp z}
\; , \label{eq:dBperp}
\end{eqnarray}
and, therefore, Eqs. (\ref{zfs1}) to (\ref{zfs3}) can be solved for $\delta \phi_z$, $\delta A_{\parallel z}$ and $\delta B_{\parallel z}$ as
independent field variables uniquely defining ZFs \cite{zonca_2021}. Note that we have maintained the last term on the RHS of Eq. (\ref{zfs1}) 
despite it usually vanishes assuming equilibrium quasineutrality. However, in the present theoretical framework where $\bar F_0$ is assumed to vary consistently with Eq. (\ref{eq:PSZSev1}),
equilibrium quasineutrality is not imposed separately, while plasma quasineutrality is satisfied overall. This means that ZFs are allowed to develop slow spatiotemporal mean field structures due to fluctuation induced transport.

PSZS, with their micro spatiotemporal scale counterpart, and the ZFs constitute the zonal state, introduced in Sec.~\ref{sec:orbitaverage}, 
which is consistent with the finite level of $n\neq 0$ symmetry breaking fluctuations. Here, the $n\neq 0$ fluctuation spectrum is 
assumed as given, but can generally be computed by means of standard nonlinear gyrokinetic theory.

\section{Zonal state self-consistent evolution}
\label{sec:org16022e2} 
In order to illuminate the evolution of the zonal state, let us focus on the case where a given $n\neq0$ spectrum is assumed.

Following~\cite{chen2016physics}, we adopt low-\(\beta\) ordering with good separation of SAW and compressional Alfvén wave frequencies and we calculate  \(\delta B_{\parallel}\) from the perpendicular pressure balance equation:
\begin{equation}
\label{eq:42}
\nabla_{\perp}\left(B_{0} \delta B_{\| z}+4 \pi \delta P_{\perp z}\right) \simeq 0 ,
\end{equation}
where \(\delta P_{\perp z}\) represents the perpendicular pressure perturbation. Having solved for \(\delta B_{\| z}\) explicitly, the fluctuation spectrum is entirely described by the scalar potential \(\delta \phi_z\) and the parallel vector potential \(\delta A_{\parallel z}\). Furthermore, for the sake of simplicity, we also assume the ZS is predominantly characterized by $\delta \phi_{z}$. Consequently, in what follows, we will describe the zonal state by means of the scalar potential ZFs only. 
Thus, we are left with the solution of the zonal component of the quasi neutrality condition, i.e., Eq.~\eqref{zfs1} and of the corresponding particle responses. The equations obtained below, may thus be readily adopted for discussing electrostatic turbulence and,  in particular, can be used to describe GAM/EGAM (energetic particle induced GAM \cite{berk_2006,nazikian_2008}) physics (cf. \ref{sec:app}). To this aim, we rewrite the gyrokinetic equation for the non adiabatic drift/banana center distribution function, i.e.:
\begin{equation}
\label{eq:45}
(\partial_{t} + v_{\parallel} \nabla_{\parallel}) \delta g_{Bz} = - e^{i Q_{z}}\left[ \frac{e}{m} \frac{\partial \bar F_{0}}{\partial \mathcal{E}} J_0\partial_{t} \delta \phi_{z} + N.L. \right]
\end{equation}
where, for the sake of brevity, the nonlinear terms \(\delta \dot{\bm X} \cdot \bm \nabla \delta F + \delta \dot{\mathcal{E}} \partial_{\mathcal{E}} \delta F\) have been indicated as \(N.L.\). Introducing the lifting of a generic scalar field to the particle phase-space~\cite{zonca2015nonlinear} and the action angle coordinates \(\vartheta_{c}\) and \(\zeta_{c}\), i.e., such that \(\omega_{b} = \dot{\vartheta}_{c}\) and \(\dot{\zeta}_{c} = \bar{\omega}_{d}\) where \(\omega_{b}\) and \(\bar\omega_{d}\) are, respectively, the bounce/transit frequency and the precession drift frequency, we obtain:
\begin{equation}
\label{eq:lifted}  
(\partial_{t} + \omega_{b} \partial_{\vartheta c})\delta g_{Bz} = - e^{i Q_{z}} \left[\frac{e}{m} \frac{\partial \bar F_{0}}{\partial \mathcal{E}} J_0 \partial_{t} \delta \phi_{z} + N.L.  \right]_{z} \; .
\end{equation}
For now, we neglect the nonlinear term and Fourier decompose the RHS with respect to the \(\vartheta_{c}\) coordinate. Meanwhile, we introduce the \(\delta \hat{G}_{l}\) function which is connected to the Fourier series of the scalar potential by the following definition:
\begin{equation}
\label{eq:46}
- e^{i Q_{z}}\frac{e}{m} \frac{\partial \bar F_{0}}{\partial \mathcal{E}} J_0 \partial_{t} \delta \phi_{z} \equiv \sum_{l} e^{i l \vartheta_{c}}\partial_{t} \delta \hat{G}_{l}. 
\end{equation}
Here, the coefficients in the Fourier series, \(\delta \hat{G}_{l}\), can be calculated as:
\begin{equation}
\delta \hat{G}_{l} \equiv \frac{1}{2 \pi} \oint d \vartheta_{c} e^{-i l \vartheta_{c}} \left[- \frac{e}{m} e^{i Q_{z}} \frac{\partial \bar F_{0}}{\partial \mathcal{E}} J_0 \delta \phi_{z} \right] = - \overline{e^{-i l \vartheta_{c} + i Q_{z}} \frac{e}{m} \frac{\partial \bar F_{0}}{\partial \mathcal{E}} J_0 \delta \phi_{z} }.
\end{equation}
It can be readily shown that the spectral representation of the linear solution of Eq.~\eqref{eq:lifted} reads:
\begin{equation}
\label{eq:50}
\delta g_{Bz} = \sum_{l} \frac{\omega_z}{\omega_z - l  \omega_{b}} e^{i l \vartheta_{c}} \delta \hat{G}_{l} \; , 
\end{equation}
where $\omega_z \equiv i \partial_t$ is the ZFs characteristic frequency that must be intended as an operator. Thus, $(\omega_z - l \omega_{b})^{-1}$ in Eq. (\ref{eq:50}) must be intended as the inverse operator of $(\omega_z - l \omega_{b})$. We emphasize that this is a formal solution since it requires integration along characteristics and, therefore, it involves an integral equation. The same procedure can be straightforwardly applied to solve the equation including the nonlinear term, and the corresponding solution can be substituted into Eq.~\eqref{zfs1}. 
Thus, restoring the species index $s$ and explicitly denoting summation on particle species as well as
summation on $\hat \sigma = \pm$, where $\hat \sigma = v_\parallel/|v_\parallel|$ for circulating particles, while, for magnetically trapped particles, $\hat \sigma = \pm$ represents the right-/left-handed rotation of the particles on the outer leg of their poloidal orbit,  the magnetic flux surface averaged Eq. (\ref{zfs1}) reads\footnote{By magnetic flux surface average we mean $[...]_\psi = (2\pi/V_\psi^\prime) \int_0^{2\pi} {\cal J} (...) d \theta$ with $V_\psi^\prime = 2\pi \int_0^{2\pi} {\cal J} d \theta$. The following equation, for simplicity, does not report a factor $(4\pi^2/V_\psi^\prime)$ that should appear in front of the double sum, $\sum_s \sum_{\hat \sigma}$, after flux surface averaging of Eq. (\ref{zfs1}).}: 

\begin{align}
&\sum_{s} \sum_{\hat \sigma} \int d \mathcal{E} d\mu \tau_{b s} \frac{e_{s}^{2}}{m_s}  \left(\overline{\frac{\partial \bar F_{0s}}{\partial \mathcal{E}} \delta \phi_{z} }
- \sum_{l} \frac{\omega_z}{\omega_z - l \omega_{b s}} \overline{e^{i l \vartheta_c -i Q_{z s}} J_0}\overline{ e^{- i l \vartheta_c + i Q_{z s}}\frac{\partial \bar F_{0s}}{\partial \mathcal{E}} J_0 \delta \phi_{z} } \right) \nonumber \\ & + \sum_{s} \sum_{\hat \sigma} \int d \mathcal{E} d\mu \frac{1}{d \psi/dr} \frac{\partial}{\partial r} \overline{\left[
\frac{e_s^{2}}{m_s} \frac{2 \mu}{\Omega_s^{2}} \frac{\partial \bar{F}_{0s}}{\partial \mu}\left(\frac{J_{0}^{2}-1}{\lambda_s^{2}}\right) \tau_{b s} \frac{d \psi}{d r} \frac{\partial}{\partial r} \delta \phi_z \right]}
 = \nonumber \\ &\sum_{s} \sum_{\hat \sigma} \int d \mathcal{E} d\mu \tau_{b s} e_{s} \sum_{l} \frac{i}{\omega_z - l \omega_{b s}} \overline{e^{i l \vartheta_c -i Q_{z s}} J_0} \, \overline{e^{ - i l \vartheta_c + iQ_{z s}}N.L.}. \label{eq:eqnforzonalstate} 
\end{align}
Note that, here, $\sum_{\hat \sigma}$ applies to circulating particles only, since it is reabsorbed by the bounce averaging for trapped particles. Furthermore, $\tau_{b s} = 2 \pi/\omega_{b s}$ and, for simplicity, we have ignored the possible contribution due to breaking the PSZS quasineutrality, discussed above. That contribution can be easily restored, 
if needed, along with the contribution of sources and collisions by letting $N.L. \rightarrow N.L. - (C_g + {\cal S})$ on the RHS of Eq. (\ref{eq:eqnforzonalstate}).
This expression has been derived by using a minimal set of assumptions that quite reasonably describe the self-consistent evolution of the ZS and, therefore, its generality make it suitable for various applications since it allows to describe an arbitrary \(\overline{F}_{0}\) while retaining realistic magnetic geometry effects. We will illustrate some of these applications in \ref{sec:app}.

In the following, we explore the low-frequency response obtained focusing on the $l=0$ component. This is clearly the response that is directly connected with transport. In particular, the hence obtained linear terms can be expressed in a compact form introducing the plasma polarizability for $s$-species, $\chi_{z s}$\footnote{ $\chi_{z s}$, is connected with the usual definition of  susceptibility, $\chi_{s}$, via the relation $\chi_{z s} = (1 + \chi_{s}) k_r^2 \lambda_D^2$, with $\lambda_D^2 = T/(4\pi n e^2)$ the Debye length.}, defined as:
\begin{eqnarray}
\hspace*{-2cm}\chi_{z s} \left[\delta \phi_z\right]_\psi & \equiv & - \frac{4\pi^2}{V_\psi^\prime} \frac{T_s}{n_sm_s} \sum_{\hat \sigma}  \int d \mathcal{E} d\mu \tau_{b s} \left\{\overline{\frac{\partial \bar F_{0s}}{\partial \mathcal{E}} \delta \phi_{z} }
- \overline{e^{-i Q_{z s}} J_0}\overline{ e^{ i Q_{z s}} J_0 \frac{\partial \bar F_{0s}}{\partial \mathcal{E}} \delta \phi_{z} } \right. \nonumber \\ & & \left. 
+ \frac{1/\tau_{bs}}{d \psi/dr} \frac{\partial}{\partial r} \overline{\left[
\frac{2 \mu}{\Omega_s^{2}} \frac{\partial \bar{F}_{0s}}{\partial \mu}\left(\frac{J_{0}^{2}-1}{\lambda_s^{2}}\right) \tau_{b s} \frac{d \psi}{d r} \frac{\partial}{\partial r} \delta \phi_z \right]} \right\} \; . \label{eq:chizs}
\end{eqnarray}
This equation becomes a closed expression for $\chi_{z s}$ once $\delta \hat \phi_z \equiv \delta \phi_z - \left[\delta \phi_z\right]_\psi $ is given\footnote{Please, note the difference between bounce, $\overline{(...)}$, and flux averaging, $[...]_\psi$, although they are the same at the lowest order for well circulating particles. The difference between $\widetilde{(...)}$ and $\widehat{(...)}$ follows consequently.}. This can be obtained from the component of the quasineutrality condition that is varying along the flux surface, and it can be shown that $|\delta \hat \phi_z| \ll |\left[\delta \phi_z\right]_\psi|$ in the long wavelength limit, $|Q_{z s}| \ll 1$ (cf., e.g., Ref. \cite{zonca_2008epl,chen_2014}). In particular, 
$\delta \hat \phi_z \rightarrow 0$ for $T_e/T_i \rightarrow 0$. Equation (\ref{eq:chizs}), valid for arbitrary wavelength, generalizes to arbitrary geometry and distribution functions the plasma polarizability expressions at short wavelengths studied recently \cite{wang__2009,lu_2019,lu_2021,cho_2021}. Introducing the  $s$-species polarization density $$[\delta n_{{\rm pol} s}]_\psi = - \frac{n_s e_s}{T_s} \chi_{z s}  \left[\delta \phi_z\right]_\psi \; , $$ the $l=0$ flux surface averaged quasineutrality condition, Eq. (\ref{eq:eqnforzonalstate}), can be rewritten as
\begin{equation}
\sum_s e_s \partial_t [\delta n_{{\rm pol} s}]_\psi = \sum_s e_s \left[ \bm \nabla \cdot \bm \Gamma_{N.L. s}\right]_\psi \; , \label{eq:transpsum}
\end{equation}
where we have introduced the flux surface averaged divergence of the $s$-species particle flux due to nonlinear interactions:
\begin{eqnarray}
\left[ \bm \nabla \cdot \bm \Gamma_{N.L. s}\right]_\psi  & = & \frac{4\pi^2}{V_\psi^\prime} \sum_{\hat \sigma} \int d \mathcal{E} d\mu \tau_{b s} \overline{e^{-i Q_{z s}} J_0} \, \overline{e^{  iQ_{z s}} N.L.}
\nonumber \\
& = & \frac{4\pi^2}{V_\psi^\prime} \frac{\partial}{\partial \psi} \sum_{\hat \sigma} \int d \mathcal{E} d\mu \tau_{b s} \left[ \overline{\left(e^{-i Q_{z s}} J_{0}\right)} \; 
\overline{\left(e^{i Q_{z s}} \delta \dot{\bar \psi}_s  \delta F_{s}\right)}  \right. \nonumber \\
& & \hspace*{5em} \left. - \overline{\frac{\partial \bar\psi_s}{\partial {\cal E}} \left(e^{-i Q_{z s}} J_{0}\right)} \; \overline{\left(e^{i Q_{z s}} \delta \dot{\cal E}  \delta F_{s}\right)} \right] \nonumber \\
& = & \frac{1}{V_{\psi}^{\prime}} \frac{\partial}{\partial \psi}\left[ V_{\psi}^{\prime} \left\langle  \left(e^{-i Q_{z s}} J_{0}\right) \; \overline{\left(e^{i Q_{z s}} \delta \dot{\bar \psi}_s  \delta F_{s}\right)}\right\rangle_{v}
\right. \nonumber \\ & & \hspace*{3em} \left. - V_{\psi}^{\prime} \left\langle  \frac{\partial \bar\psi_s}{\partial {\cal E}} \left(e^{-i Q_{z s}} J_{0}\right) \; \overline{\left(e^{i Q_{z s}} \delta \dot{\cal E}  \delta F_{s}\right)}\right\rangle_{v}
\right]_{\psi} 
\; . \label{eq:NLflux}
\end{eqnarray}
The different forms on the RHS are all equivalent and are given here to illuminate the conservation properties of these expressions.
Recalling that, at the leading order, $\delta \dot \psi = (B_0/B_\parallel^*) c \partial_\zeta \left\langle\delta L_g\right\rangle$, the last equation demonstrates that only toroidal symmetry breaking fluctuations drive a finite flux surface averaged particle transport in tokamaks at the corresponding leading order. Thus, the present analysis must assume a prescribed spectrum of $n\neq 0$ fluctuations (cf. Sec. \ref{sec:CGL}).
Physically, Eq. (\ref{eq:transpsum}) is readily interpreted as the nonlinear charge density modification compensating the polarization charge to ensure quasineutrality; that is, 
$\sum_s e_s \partial_t [\delta n_{{\rm pol} s}]_\psi = - \sum_s e_s \partial_t [\delta n_{N.L. s}]_\psi$. Meanwhile, without summing over all particle species, it is possible to cast the same equation in the form of flux surface averaged particle continuity equation:
\begin{equation} 
\partial_t \left[ n_s \right]_\psi  =  \partial_t [\delta n_{{\rm pol} s}]_\psi - \left[ \bm \nabla \cdot \bm \Gamma_{N.L. s}\right]_\psi \; , \label{eq:denscont}
\end{equation}
where collisional neoclassical transport in the banana regime as well as sources/sinks can be readily included in the expression above by letting $N.L. \rightarrow N.L. - (C_g + {\cal S})$, as discussed below Eq.  (\ref{eq:eqnforzonalstate}). Note that $\left[ n_s \right]_\psi$ on the LHS of Eq. (\ref{eq:denscont}) is the total particle density of the $s$-species, since the fluctuation induced nonlinear particle flux includes both micro- as well as meso- and macro-scale spatio-temporal behaviors, consistent with Eq. (\ref{zfs1}).
Thus, Eq. (\ref{eq:denscont}) describes the variety of spatiotemporal scales involved in particle transport. Consistently with the analysis of Ref. \cite{falessi2019}, polarization effects become important only on sufficiently short scales, $k_z L > \delta^{-1/2}$; i.e., the meso-scales, with $L$ the characteristic plasma macro-scale, $\rho_L$ the Larmor radius and $\delta = \rho_L/L$ the gyrokinetic ordering parameter.
Meanwhile, the leading order flux surface averaged particle flux for symmetry breaking fluctuations becomes
\begin{equation}
\left[ \bm \nabla \cdot \bm \Gamma_{N.L. s}\right]_\psi  = \frac{1}{V_{\psi}^{\prime}} \frac{\partial}{\partial \psi}\left[\left\langle V_{\psi}^{\prime} \left(e^{-i Q_{z s}} J_{0}\right) \; \overline{\left[c e^{i Q_{z s}} R^{2} \nabla \phi \cdot \nabla\left\langle\delta L_{g s}\right\rangle \delta g_{s}\right]}\right\rangle_{v}\right]_{\psi} \; . \label{eq:NLflux2}
\end{equation}
Again, in the $k_z L < \delta^{-1/2}$ long wavelength limit, this expression reduces to the well-known form adopted in classical analyses of fluctuation-induced evolution of macroscopic plasma profiles \cite{abel2013multiscale,balescu1988transport,shaing1988neoclassical,sugama1996transport,sugama1998nonlinear,garbet2010gyrokinetic}. Following Ref. \cite{falessi2019}, the same argument can be repeated to show that classical forms of momentum and energy transport equations are reproduced.

This demonstrates that the PSZS transport equations, Eqs. (\ref{eq:PSZSev1}) and (\ref{eq:deltagf}) derived in the previous section, along with the equations for the self-consistent determination of the ZFs, Eqs. (\ref{zfs1}) to (\ref{zfs3}), fully characterize the ZS and, at the same time, the multi-spatiotemporal-scale nature of phase-space transport in collisionless burning plasmas and their possible deviation from local thermodynamic equilibrium. This description reduces to the previous gyrokinetic theory of phase-space transport \cite{falessi2019,falessi2017gyrokineticarticle} within the framework of Frieman-Chen nonlinear gyrokinetic equation~\cite{frieman1982nonlinear} and recovers earlier works in the proper limit \cite{abel2013multiscale,balescu1988transport,shaing1988neoclassical,sugama1996transport,sugama1998nonlinear, garbet2010gyrokinetic}. Based on the nonlinear gyrokinetic theory with Hamiltonian description of particle motion accurate up ${\cal O}(\delta^2)$ \cite{sugama2017modern,brizard2007foundations}, the present PSZS transport equations are valid for even longer than the characteristic transport time scale, ${\cal O}(\delta^{-3}) \Omega^{-1}$, and typically hold on times $<{\cal O}(\delta^{-4}) \Omega^{-1}$.

\section{Conclusions and discussion}
\label{sec:concl-disc}
In this article, we have presented a comprehensive study of plasma transport processes in fusion plasmas using the phase-space zonal structure (PSZS) transport theory \cite{falessi2019,zonca_2021}.  We addressed the limitations of current numerical frameworks which are computationally expensive and often limited in their ability to capture long-time scale dynamics and non-local (global) behaviors. To overcome these challenges, we developed the PSZS transport theory, which provides a proper definition of the plasma nonlinear equilibrium distribution function by considering slowly evolving structures in the phase-space. The PSZS theory allows for the derivation of the usual plasma transport equations as a limiting case when the deviation from the local Maxwellian is small, as demonstrated in previous work \cite{falessi2019}. However, in the general case, PSZS is not associated with a reference Maxwellian, as it results from the competition between resonantly induced nonlinear transport, sources, and weakly collisional effects, necessitating a phase-space description.

Applying the PSZS transport theory, we derived the evolution equation for the zonal state (ZS), representing the renormalized nonlinear equilibrium consistent with toroidal symmetry breaking fluctuations and transport time scale ordering. Specifically, we defined the two components of the ZS, namely the PSZS and the zonal electromagnetic fields (ZFs). Moreover, applying the Chew Goldberger Low (CGL) description, we derived the self-consistent modifications of the reference magnetic equilibrium using the push forward representation of the macro-/meso-scopic component moments of the PSZS.

As an example of the theoretical framework, we discuss the self-consistent evolution of the ZS with a given spectrum of toroidally symmetry breaking perturbations and ZFs dominated by the scalar potential response. We derived expressions for the plasma polarizability that are applicable to arbitrary geometry and equilibrium distribution functions, and discussed the features of transport equations on the different spatial scales that are involved in the problem. Geodesic acoustic mode (GAM) and Energetic particle driven geodesic acoustic mode (EGAM) do not belong to the ZS due to their fast time variation and finite collisionless damping/drive. Nonetheless, finite amplitude GAM/EGAM may nonlinearly impact on the ZS evolution. Thus, we have added a detailed Appendix on this problem, where interested readers can find a discussion of the GAM/EGAM peculiar physics. In particular, we give a general expression for the linear dielectric response of GAMs. Furthermore, we illustrate examples of GAM/EGAM nonlinear dynamics, which could be readily adopted to investigate problems of practical interest in general geometry and with arbitrary energetic particles (EP) distribution functions, such as the EGAM decay into two GAMs recently observed in low-collisionality LHD plasmas \cite{Qiu_NJP_2021} and the self-consistent EGAM frequency sweeping \cite{qiu_2018}.

In conclusion, the PSZS transport theory provides a promising approach to understanding EP transport processes in fusion plasmas. The derived equations for the ZS and the associated modifications to the equilibrium provide a comprehensive framework for studying plasma nonlinear equilibrium and its evolution due to transport processes. This theoretical framework opens new possibilities for developing advanced reduced EP transport models capable of capturing the long-time scale evolution of burning plasmas and providing insight into the non-locality of transport processes. Notably, a recent advancement in this field is the proposed Dyson-Schrödinger transport Model (DSM) \cite{zonca_2021}. PSZS fluxes, computed using the \texttt{DAEPS-FALCON} suite of codes \cite{li_etal_2023, falessi2019shear}, have been calculated within the \texttt{LIGKA} EP workflow \cite{lauber_2007, popa2023ep} considering realistic Tokamak configurations. This is a crucial step towards the practical implementation of the PSZS transport theory in realistic geometry. Based on a gyrokinetic description for the underlying perturbations, employing general EP distributions functions and using saturation rules obtained from gyrokinetic non-linear codes will allow us to construct a quantitative and predictive reduced EP transport model for the interpretation of present-day experimental results and the investigation of future burning plasmas.

Future research directions include the derivation of general orbit-averaged source and collision terms on the analytical side. On the numerical side, PSZS diagnostics have been developed for global gyrokinetic and hybrid codes such as HMGC and ORB5 \cite{Bottino_2022}, enabling the study of phase-space transport processes during nonlinear gyrokinetic simulation. At present, the EP workflow calculates the PSZS evolution within the kick model \cite{podesta_2014} approximation. Further development of reduced transport models for the PSZS involves the development of a solver for the DSM \cite{zonca_2021} and the inclusion of nonlinear corrections into the governing equations of the EP workflow. More generally, a comprehensive gyrokinetic transport solver on long time scale can be developed by means of subcycling and restart of nonlinear gyrokinetic simulations in the updated ZS computed within the present theoretical framework adopting the numerically computed phase-space fluxes. These advancements will contribute to a deeper understanding of EP transport in fusion plasmas and facilitate the development of more accurate predictive models including core turbulent transport.

\ack{
This work was carried out within the framework of the EUROfusion Consortium 
and received funding from Euratom research and training programme 2014--2018 
and 2019--2020 under Grant Agreement No. 633053 (Project No. WP19-ER/ENEA-05). 
This work has been carried out within the framework of the EUROfusion Consortium, funded by the European Union via the Euratom Research and Training Programme (Grant Agreement No 101052200 -- EUROfusion; Projects No. ENR-MOD.01.MPG and AC-TSVV.10.MPG). Views and opinions expressed are however those of the author(s) only and do not necessarily reflect those of the European Union or the European Commission. Neither the European Union nor the European Commission can be held responsible for them. This work was also supported by the National Science Foundation of China Project No. 12261131622. This work was supported in part by the Italian Ministry of Foreign Affairs and International Cooperation, grant number CN23GR02 and by the MMNLP project CSN4 of INFN, Italy.
}
\appendix
\section{Applications to geodesic acoustic mode physics}
\label{sec:app}
In this appendix, as further illustration of the strength and usefulness of the present theoretical framework, we focus on selected applications of Eq. (\ref{eq:eqnforzonalstate}) derived above. Despite these results are not entirely novel, their compact derivation and validity for general geometry and distribution functions demonstrates the practical implications of the present approach. Firstly, we derive an expression for the linear dielectric response of GAM (Geodesic Acoustic Mode) oscillations, which yields the well-known results for Maxwellian distribution functions and circular equilibria \cite{fu_2008,qiu_2010} and, next, we extend the results of Ref. \cite{chen_2014} describing the generation of zero frequency ZFs by GAMs oscillations. Additionally, we investigate the modulation of GAMs by ZFs showing that the presence of energetic particles (EPs) or higher-order thermal plasma finite orbit width effects are necessary for non-vanishing nonlinear interactions as well as for second harmonic GAM generation. This generalizes the results of \cite{chen_2014,qiu_2018,zhang_2009,fu_2011}. Finally, we explore the nonlinear dynamics of EGAM highlighting the importance of PSZS and the ZS by considering the nonlinear term induced by PSZS as a nonlinear equilibrium. We describe the evolution of the zonal state, accounting for the action of sources, collisions, and the emission and re-absorption of the GAM/EGAM fluctuations. 

\subsection{Linear dielectric response of energetic particle driven geodesic acoustic mode}
\label{sec:GAMEGAMlin}

In order to calculate the linear dispersion response, we rewrite Eq. (\ref{eq:50}) taking the decomposition $\delta \phi_z = \delta \phi_G =  \left[ \delta \phi_{G} \right]_\psi + \delta \hat \phi_G$ explicitly into account, denoting that the scalar potential refers to GAM/EGAM. More precisely,
\begin{eqnarray}
\delta g_{BG} & =  &- e^{i Q_{G}}\frac{e}{m} \frac{\partial \bar F_{0}}{\partial \mathcal{E}} J_0 \delta \phi_{G} + \sum_l \left [\frac{l \omega_b  \omega_G}{\omega_G^2 - l^2 \omega_b^2} i \sin l \vartheta_c  + \frac{l^2 \omega_b^2}{\omega_G^2 - l^2 \omega_b^2} \cos l \vartheta_c \right] \nonumber \\ & & \times \left[ - \overline{e^{ i Q_{G}}  \cos l\vartheta_c  \frac{e}{m} \frac{\partial \bar F_{0}}{\partial \mathcal{E}} J_0} \right] \left[ \delta \phi_{G} \right]_\psi + \sum_l \left [\frac{l \omega_b  \omega_G}{\omega_G^2 - l^2 \omega_b^2} i \sin l \vartheta_c  \right. \nonumber \\
& & \left. + \frac{l^2 \omega_b^2}{\omega_G^2 - l^2 \omega_b^2} \cos l \vartheta_c \right] \left[ - \overline{e^{ i Q_{G}}  \cos l\vartheta_c  \frac{e}{m} \frac{\partial \bar F_{0}}{\partial \mathcal{E}} J_0 \delta \hat \phi_G} \right] \nonumber \\ & & + \sum_l \left [\frac{l \omega_b  \omega_G}{\omega_G^2 - l^2 \omega_b^2} \cos l \vartheta_c  
 + \frac{l^2 \omega_b^2}{\omega_G^2 - l^2 \omega_b^2} i \sin l \vartheta_c \right] \nonumber \\ & & \times \left[ \overline{e^{ i Q_{G}} i \sin l\vartheta_c  \frac{e}{m} \frac{\partial \bar F_{0}}{\partial \mathcal{E}} J_0 \delta \hat \phi_{G}} \right]
\; , \label{eq:dgbegam}
\end{eqnarray}
where $\delta g_{BG}$ denotes the $\delta g_{Bz}$ response to GAM, the subscript $G$ in $Q_G$ reminds that the radial shift operator acts on GAM, and we have assumed an up-down symmetric equilibrium for simplicity but without loss of generality, since the general case could be readily restored at the expense of more complicated formal expressions.
Substituting  Eq. (\ref{eq:dgbegam})  back into the linearized Eq. (\ref{zfs1}) for the varying component on the considered magnetic flux surface, we can write
\begin{eqnarray}
& & \left[ - \frac{n_e e^2}{T_e} + \sum_{s} \left\langle \frac{e_{s}^{2}}{m_s}  \frac{\partial \bar F_{0s}}{\partial \mathcal{E}}  \right\rangle_v \right] \delta \hat \phi_G 
+ \sum_s \sum_l \left\langle \left[ \frac{l \omega_{bs}  \omega_G}{\omega_G^2 - l^2 \omega_{bs}^2} i \sin l \vartheta_c J_0 e^{- i Q_{G s}} \right. \right.
 \nonumber \\ & & \hspace*{2em}  \left. \left. + \frac{\omega_G^2}{\omega_G^2 - l^2 \omega_{bs}^2} \left(\cos l \vartheta_c  J_0 e^{- i Q_{G s}} - \overline{\cos l \vartheta_c  J_0 e^{- i Q_{G s}}} \right)\right]
 \right. \nonumber \\ & & \hspace*{2em} \left. \times \left[ - \overline{e^{ i Q_{G s}}  \cos l\vartheta_c  \frac{e_s^2}{m_s} \frac{\partial \bar F_{0 s}}{\partial \mathcal{E}} J_0} \right]  \right\rangle_v 
 \left[ \delta \phi_{G} \right]_\psi + \sum_s \sum_l \left\langle \left[ \frac{l \omega_{bs}  \omega_G}{\omega_G^2 - l^2 \omega_{bs}^2} \right. \right. \nonumber  \\
& & \hspace*{2em} \times \left.  i \sin l \vartheta_c  J_0 e^{- i Q_{G s}}  
 + \frac{\omega_G^2}{\omega_G^2 - l^2 \omega_{bs}^2} \left(\cos l \vartheta_c  J_0 e^{- i Q_{G s}} - \overline{\cos l \vartheta_c  J_0 e^{- i Q_{G s}}} \right) \right]
 \nonumber \\ & & \hspace*{2em} \left.  \times \left[ \overline{ - e^{ i Q_{G s}}  \cos l\vartheta_c  \frac{e_s^2}{m_s} \frac{\partial \bar F_{0 s}}{\partial \mathcal{E}} J_0 \delta \hat \phi_G} \right]  \right\rangle_v 
 + \sum_s \sum_l \left\langle \left[ \frac{l \omega_{bs}  \omega_G}{\omega_G^2 - l^2 \omega_{bs}^2} \right. \right. \nonumber  \\
& & \hspace*{2em} \times \left. \left(\cos l \vartheta_c  J_0 e^{- i Q_{G s}} - \overline{\cos l \vartheta_c  J_0 e^{- i Q_{G s}}} \right)  
 + \frac{\omega_G^2}{\omega_G^2 - l^2 \omega_{bs}^2} i \sin l \vartheta_c  J_0 e^{- i Q_{G s}} \right]
 \nonumber \\ & & \hspace*{2em} \left.  \times \left[ \overline{e^{ i Q_{G s}}  i \sin l\vartheta_c  \frac{e_s^2}{m_s} \frac{\partial \bar F_{0 s}}{\partial \mathcal{E}} J_0 \delta \hat \phi_G} \right]  \right\rangle_v  = 0\; , \label{eq:dphihat}
\end{eqnarray}
where, for simplicity, we have assumed Maxwellian electrons. Equation (\ref{eq:dphihat}) is readily solved for $\delta \hat \phi_G$ as a function of $ \left[ \delta \phi_{G} \right]_\psi $ reducing to 
well known results, e.g. \cite{gao_2006,sugama2006collisionless,zonca_2008epl}, for Maxwellian ions in the long wavelength limit. Meanwhile, the flux surface averaged quasineutrality condition can be written as
\begin{eqnarray}
& &  \hspace*{-2em}  \frac{1}{V_\psi'}\frac{1}{d\psi/dr} \frac{\partial}{\partial r} \left[ V_\psi'  \sum_s \rho_{L s}^2 \frac{n_s e_s^2}{T_s} D_{Gs} \frac{d \psi}{dr}
 \frac{\partial}{\partial r} \left[ \delta \phi_{G} \right]_\psi  \right] 
=  \frac{4\pi^2}{V_\psi'} \sum_{s} \sum_{\hat \sigma} \int d \mathcal{E} d\mu \tau_{b s} e_{s} \nonumber \\
& &\times \sum_{l} \frac{i}{\omega_G - l \omega_{b s}} \overline{\cos l \vartheta_c  J_0 e^{- i Q_{G s}}} \, \overline{e^{ - i l \vartheta_c + iQ_{G s}}N.L.} \; , \label{eq:GAMqn}
\end{eqnarray}
where $\rho_{L s}^2 = (T_s/m_s)/\bar \Omega_s^2$, the temperature $T_s$ is defined as $T_s \equiv n_s^{-1} \left\langle 2 m_s \mu B_0 \bar F_{0s}  \right\rangle_v$ for a generic non-Maxwellian distribution function, $\bar \Omega_s$ is the cyclotron frequency computed at the on-magnetic-axis magnetic field $B_0 = \bar B_0$, and $D_{Gs}$ is the $s$-species contribution to the GAM/EGAM dispersion response, expressed as, noting Eq.  (\ref{eq:dphihat}):
\begin{eqnarray}
& &  \hspace*{-2em} D_{Gs} \left[ \delta \phi_{G} \right]_\psi = \frac{4\pi^2}{V_\psi'} \sum_{\hat \sigma} \int d \mathcal{E} d\mu \tau_{b s} \left\{ \left[ \overline{\left[ 
\frac{2 \mu \bar B_0^2}{n_s B_0} \left(\frac{J_{0}^{2}-1}{\lambda_s^{2}}\right)  \left(  \frac{\partial \bar{F}_{0s}}{\partial {\cal E}} +  \frac{1}{B_0} \frac{\partial \bar{F}_{0s}}{\partial \mu} \right) \right]} 
 \right. \right. \nonumber \\ & &  \left. + \sum_l \frac{l^2 \omega_{bs}^2}{\omega_G^2 - l^2 \omega_{bs}^2}  \overline{\cos l \vartheta_c  J_0 e^{- i Q_{G s}}} \overline{ \left[ e^{ i Q_{Gs}}  \cos l\vartheta_c   \frac{\bar \Omega_s^2}{n_s k_r^2} \frac{\partial \bar F_{0s }}{\partial \mathcal{E}} J_0 \right]} \right] 
 \left[ \delta \phi_{G} \right]_\psi \nonumber \\ & &
\left. + \sum_l \frac{\omega_G^2}{\omega_G^2 - l^2 \omega_{bs}^2}  \overline{\cos l \vartheta_c  J_0 e^{- i Q_{G s}}} \overline{ \left[ e^{ i Q_{Gs}}  \cos l\vartheta_c  \frac{\bar \Omega_s^2}{n_s k_r^2} \frac{\partial \bar F_{0s }}{\partial \mathcal{E}} J_0 \delta \hat \phi_G\right]} \right]  \nonumber \\ & &
\left. \left. - \sum_l \frac{l \omega_{bs}\omega_G}{\omega_G^2 - l^2 \omega_{bs}^2}  \overline{\cos l \vartheta_c  J_0 e^{- i Q_{G s}}} \overline{ \left[ e^{ i Q_{G s}}  i \sin l\vartheta_c  \frac{\bar \Omega_s^2}{n_s k_r^2} \frac{\partial \bar F_{0s }}{\partial \mathcal{E}} J_0 \delta \hat \phi_G\right]} \right]  \right\}
  \; . \label{eq:DGs}
\end{eqnarray}
Note that electrons do not give contribution the GAM/EGAM dispersion response since they cannot respond to $n=0$ perturbations in the GAM/EGAM frequency range, as it is well known. Equation (\ref{eq:DGs}) generalizes previously derived expressions of the GAM/EGAM dispersion relation \cite{fu_2008,qiu_2010} (cf. Ref. \cite{qiu_2018} for a recent review) to the case of general geometry and distribution functions, and recovers them in the proper limit; e.g., for circular cross section tokamak equilibria, where, upon expanding $e^{ i Q_{Gs}} \simeq 1 +  i Q_{Gs}$ in the long wavelength limit,
\begin{equation}
i Q_{Gs} \simeq \left[ 1 + \left( 1 + \frac{\mu \bar B_0}{\bar v_\parallel^2} \right) \frac{r}{R_0} \cos \theta \right] \frac{q R_0}{r} \frac{\bar v_\parallel}{\bar \Omega_s} \partial_r \; , \label{eq:QGs}
\end{equation}
with $R = R_0$ denoting the magnetic axis, $\bar v_\parallel$ the parallel velocity at $\bar B_0$, and 
we also have $V_\psi' = 4 \pi^2 q R_0/\bar B_0$ and $\tau_{b} = 2 \pi q R_0/|\bar v_\parallel|$ for well circulating particles. In fact, assuming $|\omega_G|\gg \omega_{b}$ and one single ion species, Eq.  (\ref{eq:dphihat}) yields
$$\delta \hat \phi_G \simeq 2 \frac{T_e}{T_i} \frac{T_i/m_i}{\bar \Omega_i \omega_G} \frac{i}{R_0} \sin \theta \frac{\partial}{\partial r}  \left[ \delta \phi_{G} \right]_\psi  \; , $$
having noted that $\vartheta_c \simeq \hat \sigma \theta$ for well circulating particles, while Eq. (\ref{eq:DGs}) reduces to
$$D_{Gi} \simeq 1 - \frac{2T_i/m_i}{R_0^2 \omega_G^2} \left( \frac{7}{4} + \frac{T_e}{T_i} \right) \;,$$
from which the leading order GAM frequency can be obtained. The GAM collisionless damping  and/or resonant EGAM excitation by phase-space anisotropic EPs can be obtained from the wave-particle resonances embedded in Eq. (\ref{eq:DGs}). Meanwhile, GAM/EGAM collisional damping is readily restored by letting $N.L. \rightarrow N.L. - (C_g + {\cal S})$ on the RHS of Eq. (\ref{eq:GAMqn}) (cf. Eq.  (\ref{eq:eqnforzonalstate}) in Sec.  \ref{sec:org16022e2}). Most importantly, however, the formally nonlinear term on the RHS of Eq. (\ref{eq:GAMqn}) allows us to discuss the relative role of different processes contributing to GAM/EGAM nonlinear dynamics  \cite{chen_2014,qiu_2018}, which are addressed in the next two subsections.

\subsection{Zero frequency zonal flow generation by geodesic acoustic modes}
\label{sec:ZFZFbyGAM}

Consider the generation of zero frequency zonal flow by self-modulation of interacting GAMs. In particular, let us look at the flux surface averaged quasineutrality condition, Eq. (\ref{eq:eqnforzonalstate}), in the form of Eq. (\ref{eq:transpsum}); i.e.,
\begin{equation}
\sum_s \frac{n_s e^2_s}{T_s} \chi_{z s}  \left[\delta \phi_z\right]_\psi  = - \sum_s e_s \partial_t^{-1} \left[ \bm \nabla \cdot \bm \Gamma_{N.L. s}\right]_\psi \; , \label{eq:zfzf0}
\end{equation}
where $\chi_{z s}$ is the general polarizability expression derived above in Eq. (\ref{eq:chizs}). In order to calculate the nonlinear flux due to GAM, we assume that mode frequency is much larger than bounce/transit frequency, $|\omega_G| \gg \omega_b$, and, thus, from Eq. (\ref{eq:dgbegam}),
\begin{equation}
\delta g_{BG} \simeq - e^{i Q_{G}}\frac{e}{m} \frac{\partial \bar F_{0}}{\partial \mathcal{E}} J_0 \delta \phi_{G} + i \sum_l \frac{l\omega_b}{\omega_G} \sin l \vartheta_c  \left[ - \overline{e^{ i Q_{G}}   \cos l\vartheta_c  \frac{e}{m} \frac{\partial \bar F_{0}}{\partial \mathcal{E}} J_0} \right] \left[ \delta \phi_{G}  \right]_\psi\; , \label{eq:dgb}
\end{equation}
up to first order in the $\omega_b/\omega_G$ expansion. Note that all bounce harmonics are retained due to the fact that GAM are characterized by finite frequency and that, similarly to Eq. (\ref{eq:dgbegam}),  
we have considered an up-down symmetric equilibrium for simplicity but without loss of generality.
Now, let's note that $\delta \hat \phi_G = {\cal O}(k_z \rho_L) [ \delta \phi_G]_\psi$ for GAM  \cite{zonca_2008epl,chen_2014} and, thus, that linear as well as nonlinear dynamics are dominated by finite orbit width effects. As a consequence, the nonlinear flux due to GAM on the RHS of Eq. (\ref{eq:zfzf0}) is dominated by the $\propto e^{i Q_{z}} \dot \theta_z \partial_\theta \delta F_z$ term.
Noting also that
$$ \delta \dot \theta_z = - \frac{c R B_\phi}{{\cal J} B_0 B_\parallel^* (d\psi/d r)} (J_0 \delta E_{r z}) \simeq - \frac{c R B_\phi}{{\cal J} B_0^2  (d\psi/d r)} (J_0 \delta E_{r z}) $$ 
at the leading order, where $\delta E_{r z}$ is the GAM radial electric field, we have 
\begin{equation}\delta E_{r z} \simeq \frac{1}{2} \left( \delta E_{r G} (r, t) e^{-i \omega_G t} + \delta E_{r G}^* (r, t) e^{i \omega_G t} \right) \; ; \label{eq:dErG}
\end{equation}
and, thus, 
\begin{eqnarray}
\overline{e^{  iQ_{z}}N.L.} & = &  \left[  \hat \sigma \sum_l \overline{e^{ i Q_{G}}  \cos l\vartheta_c  \frac{c R B_\phi}{4 {\cal J} B_0^2  (d\psi/d r)} (J_0 \delta E_{r G})^* }  \right. \nonumber \\
& & \times \left. \frac{i l^2\omega_b}{(\omega_G + i\partial_t)} \overline{e^{ i Q_{G}}  \cos l\vartheta_c  \frac{e}{m} \frac{\partial \bar F_{0}}{\partial \mathcal{E}} J_0 \delta \phi_{G} } + c.c. \right] \nonumber \\
& \simeq & \partial_t \sum_l \frac{ l^2 \hat \sigma\omega_b}{\omega_G^2} \overline{e^{ i Q_{G}}  \cos l\vartheta_c  \frac{c R B_\phi}{4 {\cal J} B_0^2  (d\psi/d r)} (J_0 \delta E_{r G})^* }  \nonumber \\
& & \times  \overline{e^{ i Q_{G}}  \cos l\vartheta_c  \frac{e}{m} \frac{\partial \bar F_{0}}{\partial \mathcal{E}} J_0 \delta \phi_{G} }  + c.c. \; , 
\end{eqnarray}
where we have assumed that $\partial_\theta \sin l \vartheta_c \simeq l \hat \sigma \cos l \vartheta_c$ for well circulating particles \cite{zonca2015nonlinear}\footnote{As for the case of up-down symmetric equilibria, this 
assumption simplifies notations but can be generally relaxed when carrying out numerical quadratures, which allow using the general map $\theta \mapsto \vartheta_c$ for given constants of motion $(P_\phi, {\cal E}, \mu)$.}, $c.c.$ stands for complex conjugate and $(\omega_G + i\partial_t)^{-1}$ is the inverse of $(\omega_G + i\partial_t)$. Equation (\ref{eq:zfzf0}), thus, becomes
\begin{eqnarray}
\sum_s \frac{n_s e^2_s}{T_s} \chi_{z s}  \left[\delta \phi_z\right]_\psi & = & \frac{4\pi^2}{V_\psi^\prime} \frac{\partial}{\partial \psi}  \sum_s \sum_{\hat \sigma} \int d \mathcal{E} d\mu \tau_{b s} \overline{e^{-i Q_{z s}} J_0} 
\nonumber \\ & & \times \sum_l \frac{ l^2 \hat \sigma\omega_{b s}}{\omega_G^2} \overline{e^{ i Q_{Gs}}  \cos l\vartheta_c  \frac{c R B_\phi}{4 {\cal J} B_0^2} (J_0 \phi_{G})^* }
\nonumber \\ & & \times  \overline{e^{ i Q_{Gs}}  \cos l\vartheta_c  \frac{e^2_s}{m_s} \frac{\partial \bar F_{0 s}}{\partial \mathcal{E}} J_0 \delta \phi_{G} }  \; .\label{eq:zfzf1}
\end{eqnarray}
This expression generalizes that derived in Ref. \cite{chen_2014} and reduces to it upon expanding $e^{ i Q_{Gs}} \simeq 1 +  i Q_{Gs}$ in the long wavelength limit and noting Eq. (\ref{eq:QGs})
for a high aspect-ratio tokamak equilibrium. Consistent with  \cite{chen_2014}, Eq. (\ref{eq:zfzf1}) suggests that efficient generation of zero frequency zonal flow by GAM occurs at short wavelength due to finite orbit width effects. Meanwhile, in the long wavelength limit, the leading order response is finite only for distribution functions that are not even in $\hat \sigma$. For distribution functions that are symmetric in $\hat \sigma$, retaining higher order contributions in the  $e^{ i Q_{Gs}}$ and $e^{ - i Q_{z s}}$ expansions is necessary for computing the leading order non-vanishing term on the RHS of 
Eq. (\ref{eq:zfzf1}), as it was shown in Ref. \cite{ren_2020} for the case of a bi-Maxwellian $\bar F_0$, which is readily recovered from Eq. (\ref{eq:zfzf1}) in the proper limit.

\subsection{Null modification of GAM by ZFs nor GAM second harmonic generation}
\label{sec:null}

Let us first consider the GAM modulation by the ZFs generated either the process discussed in \ref{sec:ZFZFbyGAM} or by $n\neq 0$ toroidal symmetry breaking fluctuations. In Eq. (\ref{eq:GAMqn}) with $|\omega_G| \gg \omega_b$, the relevant nonlinear term reads
\begin{eqnarray}
& & \frac{i}{\omega_{G}} \int d{\cal E}  \tau_{bs} \overline{J_0 N.L.} = \frac{i}{\omega_{G}} \int d{\cal E} \oint d \theta J_0 \left\{ \left[ \frac{c RB_\phi}{B_0 v_\parallel} \frac{\partial}{\partial \psi} \left( J_0 \left[
\delta \phi_z \right]_\psi\right) \frac{\partial}{\partial \theta} \right. \right.\nonumber \\
& & \hspace*{2em} \left. -  \frac{\partial}{\partial \theta}\left( \frac{c RB_\phi v_\parallel}{B_0} \right) \frac{\partial}{\partial \psi} \left( J_0 \left[
\delta \phi_z \right]_\psi \right) \frac{\partial}{\partial {\cal E}} \right] \delta F_{Gs} \nonumber \\ & & \hspace*{2em} + \left[ \frac{c RB_\phi}{B_0 v_\parallel} \frac{\partial}{\partial \psi} \left( J_0 \left[
\delta \phi_G \right]_\psi\right) \frac{\partial}{\partial \theta} -  \frac{\partial}{\partial \theta}\left( \frac{c RB_\phi v_\parallel}{B_0} \right) \right.
\nonumber \\ & &  \hspace*{2em} \left. \left. \times \frac{\partial}{\partial \psi} \left( J_0 \left[
\delta \phi_G \right]_\psi \right) \frac{\partial}{\partial {\cal E}} \right] \delta F_{zs} \right\}  \; , \label{eq:GAMZFS}
\end{eqnarray}
at the leading order in the ${\cal O}(k_z q\rho_L), {\cal O}(k_G \rho_L)$ expansion, where $\delta F_G$ denotes the GAM particle response, while $\delta F_z$ is the low frequency particle response consistent with the ZFs generated nonlinearly in Eq. (\ref{eq:zfzf1}). Here, the first terms in the square brackets represent $\delta \dot \theta_{z,G} \partial_\theta$, respectively, while the second ones stand for $\delta \dot {\cal E}_{z,G} \partial_{\cal E}$.  Integrating by parts in $\theta$ the first terms and by parts in ${\cal E}$ the second ones, it can be recognized that this expressions vanish at the leading order, which means that GAM cannot  be modulated by ZFs, either generated by self-modulation or by other by $n\neq 0$ toroidal symmetry breaking fluctuations on the parallel nonlinearity time scale. This result is consistent with the findings of Ref. \cite{chen_2014}. 

Let us  now reconsider the GAM self-modulation and compute the generation of GAM second harmonic. Equation (\ref{eq:GAMZFS}) with $|\omega_G| \gg \omega_b$ can be specialized to this case and becomes, at the leading order in the ${\cal O}(k_G \rho_L)$ expansion,
\begin{eqnarray}
& & \frac{i}{\omega_{GII}} \int d{\cal E}  \tau_{bs} \overline{J_0 N.L.} = \frac{i}{\omega_{GII}} \int d{\cal E} \oint d \theta J_0 \left[ \frac{c RB_\phi}{B_0 v_\parallel} \frac{\partial}{\partial \psi} \left( J_0 \left[
\delta \phi_G \right]_\psi\right) \frac{\partial}{\partial \theta} \right. \nonumber \\
& & \hspace*{2em} \left. -  \frac{\partial}{\partial \theta}\left( \frac{c RB_\phi v_\parallel}{B_0} \right) \frac{\partial}{\partial \psi} \left( J_0 \left[
\delta \phi_G \right]_\psi \right) \frac{\partial}{\partial {\cal E}} \right] \delta F_{Gs}\; . \label{eq:GAMII}
\end{eqnarray}
Here, $\omega_{GII} \simeq 2 \omega_G$ stands for GAM second harmonic possibly driven by the considered finite amplitude GAM.
Again, integrating by parts in $\theta$ the first term in square brackets and by parts in ${\cal E}$ the second one,  this expression vanishes at the leading order, which means that GAM self-modulation cannot generate second harmonic GAM on the parallel nonlinearity time scale. Second harmonic GAM generation becomes possible by inclusion of EP nonlinear dynamics in the GAM self-modulation or higher order thermal plasma finite orbit width effects. Equation (\ref{eq:GAMII}) generalizes to shaped geometry and  arbitrary distribution functions the original result of Ref. \cite{chen_2014,qiu_2018,zhang_2009,fu_2011}.

\subsection{Nonlinear dynamics of energetic particle driven geodesic acoustic modes}
\label{sec:NLEGAM}

When looking at GAM excited by EPs, the assumption $|\omega_G| \gg \omega_b$ underlying the derivations in \ref{sec:null} is not applicable any longer and that bears consequences for the GAM-ZFs and GAM-PSZS interactions. Let us reconsider Eq. (\ref{eq:GAMqn}) with the most general nonlinear interaction term on the RHS. We analyze first the PSZS induced nonlinear term, noting that, in the low-frequency limit, 
$$\delta F_z = e^{-i Q_z} \overline{e^{iQ_z} \delta F_z} \; . $$
This means that the corresponding low-frequency $\delta F_z$ is a function only of $(P_\phi, {\cal E}, \mu)$.  Thus, when looking at its nonlinear interaction with a generic fluctuation structure, including the $n=0$ GAM/EGAM, we have
\begin{eqnarray}
& & \left( \delta \dot \psi_G \partial_\psi + \delta \dot \theta_G \partial_\theta + \delta \dot{\cal E}_G \partial_{\cal E} \right) \delta F_z = - \left( \partial_t + \dot{\bm X}_0 \cdot \bm \nabla \right) \left( \frac{e}{m} \left\langle \delta L_{gG} \right\rangle \right) \frac{\partial \delta F_z}{\partial {\cal E}}  \nonumber \\
& & \hspace*{2em} 
+ \left( \partial_t + \dot{\bm X}_0 \cdot \bm \nabla \right) \left( \frac{RB_\phi \left\langle \delta A_{\parallel g G} \right\rangle}{B_0} \right) \frac{\partial \delta F_z}{\partial \bar \psi} 
+ \frac{e}{m} \partial_t \left \langle \delta L_{gG} \right\rangle \frac{\partial \delta F_z}{\partial {\cal E}}  \; , \label{eq:PSZSrenorm0}
\end{eqnarray}
where we have noted Eqs. (\ref{eq:3}) and (\ref{eq:4}). Now recall Eq. (\ref{eq:9}) along with Eqs. (\ref{eq:45}) and (\ref{eq:lifted}). Thus, when computing the contribution of the first term on the RHS 
above to the nonlinear interaction term in Eq. (\ref{eq:GAMqn}), we have
\begin{equation}
\overline{e^{ - i l \vartheta_c + iQ_{G}}N.L.} = i (\omega_G - l \omega_b) \overline{e^{ - i l \vartheta_c + iQ_{G }} \left( \frac{e}{m} \left\langle \delta L_{g G} \right\rangle \right)}
 \frac{\partial}{\partial {\cal E}}  \overline{e^{iQ_z} \delta F_z}\; .  \label{eq:PSZSrenorm1}\end{equation}
A similar equation can be derived for the nonlinear contribution of the second term on the RHS in Eq. (\ref{eq:PSZSrenorm0}). As a consequence, we can incorporate the low frequency response into the zonal state, by allowing a fast spatial variation, consistent with Eq. (\ref{eq:F0*}) as well as Eqs. (\ref{eq:31}) and (\ref{eq:11}), preserving the structure of the governing equations. Physically, this means that the low frequency distortion in the particle distribution function can be treated as a nonlinear equilibrium and corresponds to the renormalization of particle response discussed in Sec. \ref{sec:renorm}. It also further illuminates the physical meaning of PSZS and zonal state.

We analyze now the effect of ZFs on GAM/EGAM. At the leading order, we have
\begin{equation}
e^{iQ_G}  \left( \delta \dot \psi_z \partial_\psi + \delta \dot \theta_z \partial_\theta + \delta \dot{\cal E}_z \partial_{\cal E} \right) e^{-iQ_G}  \delta F_{BG} 
= \left[ e^{iQ_z}  \left( \delta \dot \theta_z \partial_\theta + \delta \dot{\cal E}_z \partial_{\cal E} \right)\right]  \delta F_{BG}  \; . \label{eq:sr0}
\end{equation}
Therefore, the propagator $(\omega_z - l \omega_b)^{-1}$ in Eq. (\ref{eq:50}) is renormalized as
\begin{equation}
\left( \omega_G + i \partial_t - l \omega_b - \Delta_1 \right)^{-1} \; ,
\label{eq:sr1} \end{equation}
where 
\begin{equation}
\Delta_1 = - i  \overline{e^{-i l \vartheta_c} \left[ e^{iQ_z}  \left( \delta \dot \theta_z \partial_\theta + \delta \dot{\cal E}_z \partial_{\cal E} \right)\right]  e^{i l \vartheta_c}} \; .
\label{eq:ZFSdec} \end{equation}
Here, $\{ ... \}^{-1}$ denotes the inverse operator and, for simplicity, we have assumed isolated resonances, nearby which, given the results of Sec. \ref{sec:null}, we can expect that the dominant nonlinear effects occur. The overlapping resonance case can be handled by a similar approach at the price of additional technical complications. Note that Eq. (\ref{eq:ZFSdec}) describes the ``shearing'' effect of the ZFs, that is the wave particle decorrelation effect near resonance due to the poloidal flow as well as the low-frequency axisymmetric energy redistribution due to ZFs. The latter is typically negligible for toroidal symmetry breaking fluctuations but, in general, needs to be taken into account for a proper treatment of the effect of $n=0$ GAM/EGAM. Following a similar argument and Dupree's classical approach to resonance broadening theory \cite{dupree1966perturbation}, we can also evaluate the effect of renormalization of the propagator in Eq. (\ref{eq:sr1}) by generation of second harmonic component in the particle distribution functions. Noting that, 
near the $\omega_G = l \omega_b$ resonance, where $\delta F_{B G} \simeq e^{i l \vartheta_c} \delta F_{B G}^{(l)}$,
\begin{equation}
\delta F_{B GII} \simeq - i \sum_{l'} \frac{e^{i l' \vartheta_c}}{(\omega_{G II} - l' \omega_b)} \overline{e^{-i l' \vartheta_c} \left[ e^{iQ_G}  \left( \delta \dot \theta_G \partial_\theta + \delta \dot{\cal E}_G \partial_{\cal E} \right)\right]  e^{i l \vartheta_c}} \, \delta F_{B G}^{(l)} \; ,
\label{eq:dFGBII} \end{equation}
where, noting Eq. (\ref{eq:dErG}),
\begin{eqnarray*}
\delta \dot \theta_G & = & -  \frac{c RB_\phi}{{\cal J} B_0 B_\parallel^*} \frac{J_0 \delta E_{rG}}{2 d\psi/dr} \, , \\
\delta \dot {\cal E}_G & = & \frac{c v_\parallel}{{\cal J} B_\parallel^*} \frac{\partial}{\partial \theta} \left( \frac{RB_\phi v_\parallel}{B_0} \right) \frac{J_0 \delta E_{rG}}{2 d\psi/dr} \, .
\end{eqnarray*}
Thus, the further renormalization of the propagator in Eq. (\ref{eq:sr1}) yields
\begin{equation}
\left( \omega_G + i \partial_t - l \omega_b - \Delta_1  - \Delta_2 \right)^{-1} \; ,
\label{eq:sr2} \end{equation}
with
\begin{eqnarray}
\Delta_2 & = &- \sum_{l'}  \overline{e^{-i l \vartheta_c} \left[ e^{iQ_G}  \left( \delta \dot \theta_G \partial_\theta + \delta \dot{\cal E}_G \partial_{\cal E} \right)^*\right]  e^{i l' \vartheta_c}}
(\omega_{GII} + i \partial_t - l' \omega_b - \Delta_1 )^{-1} \nonumber \\ & & \times
\overline{e^{-i l' \vartheta_c} \left[ e^{iQ_G}  \left( \delta \dot \theta_G \partial_\theta + \delta \dot{\cal E}_G \partial_{\cal E} \right)\right]  e^{i l \vartheta_c}}  \; ,
\label{eq:broadgen} \end{eqnarray}
where, for completeness, we have added long time scale dependences in the propagator together with the effect of ZFs on wave-particle decorrelation.
Equation (\ref{eq:broadgen}) accounts for nonlinear frequency shift as well as of resonance broadening \cite{dupree1966perturbation}, acting as spontaneous nonlinear regulation of the minimum resonance width for a coherent nearly-periodic spectrum \cite{zoncarmpp2021}. In summary, the GAM/EGAM nonlinear problem is formally linear and given by Eq. (\ref{eq:GAMqn}) with vanishing RHS, where, however, in Eq. (\ref{eq:DGs}) the PSZS is 
given by Eq. (\ref{eq:F0*}) and the renormalized propagator by Eq. (\ref{eq:sr2}). This is consistent with the findings of \ref{sec:null}, predicting null nonlinear interactions in the GAM-ZFs and GAM-GAM system when wave-particle interactions are neglected  \cite{chen_2014}. The nonlinear system is closed by the PSZS evolution equation, Eqs. (\ref{eq:PSZSev1}) and (\ref{eq:deltagf}), which, near the $\omega_G \simeq l \omega_b$ resonance, can be combined as:
\begin{eqnarray}
& & \hspace*{-4em} \partial_t \overline{e^{i Q_z} F_{0*}}  = \left. \overline{e^{i Q_z} \left[ C_g + {\cal S} \right]} \right|_{z} + \frac{1}{2} \left[\overline{e^{i Q_G} \cos l\vartheta_c
\frac{e}{m} J_0 |\delta \phi_G| }\right]
\label{eq:F0*evol} \\ & & \hspace*{-2em} \times \frac{\partial}{\partial {\cal E}} \left\{l^2 \omega_b^2 \left[  (\omega_G - l \omega_b)^2 + \partial_t^2 \right]^{-1}  
\partial_t  \left[ \overline{e^{ i Q_{G}}  \cos l\vartheta_c  \frac{e}{m} \frac{\partial \bar F_{0*}}{\partial \mathcal{E}} J_0 \left|\delta \phi_{G} \right| } \right] \right\} 
\; , \nonumber
\end{eqnarray}
where integration by parts was made to obtain the first $\overline{[...]}$ on the RHS. 
Here, for simplicity, we have dropped $\Delta_1$ and $\Delta_2$ terms in Eq. (\ref{eq:sr2}) and noted the result of \ref{sec:ZFZFbyGAM} to neglect the $\propto \dot \theta_G \partial_\theta$ contribution to the nonlinear response for $F_{0*}$ symmetric in $\hat \sigma$. We have also assumed $T_e/T_i \ll 1$, without loss of generality, in order to drop $\delta \hat \phi_G$ with respect to $[ \delta \phi_G]_\psi$ following Ref. \cite{qiu_2018}. Equation (\ref{eq:F0*evol}) represents the evolution of the zonal state under the action of sources and collisions, as well as of emission and re-absorption of the GAM/EGAM fluctuations. In this respect, neglecting sources and collisions, Eq. (\ref{eq:F0*evol}) is a Dyson-like equation \cite{dyson49,schwinger51} as noted earlier \cite{zonca2015nonlinear,chen2016physics,zonca2014energetic,zonca_2021}, and its solution, which can be formally represented as a Dyson series, describes the evolution of the ZS. 
Equation (\ref{eq:F0*evol}) is given in time representation and is the extension to general geometry of the analogous equation for the evolution of the renormalized fast ion distribution function given in Ref. \cite{qiu_2018} using the frequency representation. Equations  (\ref{eq:GAMqn})  and (\ref{eq:F0*evol}) are perhaps one of the simplest possible illustrations of the DSM to the self-consistent evolution of the ZS.  More detailed analyses of Eqs.  (\ref{eq:GAMqn})  and (\ref{eq:F0*evol}) are beyond the present scope of illustration of simple applications of the general theoretical framework; thus, they will be reported elsewhere.

\printbibliography

\end{document}